\documentclass[prb,preprint,showpacs]{revtex4}
\usepackage{amssymb}
\usepackage{epsfig}
\usepackage{graphicx}

\begin{document}

\title{\textbf{Kinetic Theory for Electron Dynamics Near a Positive Ion}}
\author{Jeffrey M. Wrighton and James W. Dufty}
\affiliation{Department of Physics, University of Florida, Gainesville, FL 32611}
\date{\today }

\begin{abstract}
A theoretical description of time correlation functions for electron
properties in the presence of a positive ion of charge number $Z$ is given.
The simplest case of an electron gas distorted by a single ion is
considered. A semi-classical representation with a regularized electron -
ion potential is used to obtain a linear kinetic theory that is
asymptotically exact at short times. This Markovian approximation includes
all initial (equilibrium) electron - electron and electron - ion
correlations through renormalized pair potentials. The kinetic theory is
solved in terms of single particle trajectories of the electron - ion
potential and a dielectric function for the inhomogeneous electron gas. The
results are illustrated by a calculation of the autocorrelation function for
the electron field at the ion. The dependence on charge number $Z$ is shown
to be dominated by the bound states of the effective electron - ion
potential. On this basis, a very simple practical representation of the
trajectories is proposed and shown to be accurate over a wide range
including strong electron - ion coupling. This simple representation is then
used for a brief analysis of the dielectric function for the inhomogeneous
electron gas.
\end{abstract}

\date{\today }
\maketitle

%\draft

\section{Introduction}

\label{sec1}

Electron dynamics in a rigid uniform neutralizing background is a
well-studied problem (jellium in quantum mechanics \cite{fetter}, one
component plasma in classical mechanics \cite{hansen}). More realistically,
point ions lead to a polarization of the electron density (e.g., in a
hydrogen plasma) and the dynamics of the non-uniformly distributed electrons
is radically changed. The objective here is to provide a practical theory
for the description of equilibrium time correlation functions for electrons
in the simplest case of a single point ion of charge number $Z$. If the
charge is positive, essential quantum diffraction effects must be accounted
for even at high temperatures and low densities to avoid the electron - ion
Coulomb singularity. A classical Hamiltonian description is used here, with
a regularized electron - ion interaction that accounts for such effects \cite%
{filinov}. This study is an outgrowth of recent investigations based on
molecular dynamics simulations for this system \cite{ilya}. The qualitative
features observed for the electron field autocorrelation function from
simulation were captured by a simple mean field kinetic theory. Such a
kinetic theory is obtained here from the asymptotically exact short time
limit for the generator of the dynamics, providing both context and a
generalization of the analysis in reference 4 to strong electron - electron
coupling conditions.

The kinetic theory is solved exactly to express the correlation functions in
terms of effective single electron trajectories about the ion and collective
excitations via a dielectric function for the non-uniform electron fluid.
For $Z=0$ the results reduce to the familiar random phase approximation
(RPA) with local field effects (the generalized Vlasov approximation of
reference 5; see also reference 6 for a related nonlinear kinetic equation
for dusty plasmas). More generally it constitutes a generalization of the
RPA to a non-uniform electron gas, with both ion - electron and electron -
electron interactions renormalized by correlations (the Vlasov equation for
an electron gas in a periodic potential is discussed in reference 7). The
only required input is the time independent correlations for one or two
electrons and the ion. For the calculations here the hypernetted chain
approximation (HNC) integral equations are used for these static
correlations. The correlation functions are further decomposed into
contributions from the bound and free (positive and negative energy) states
of the effective single particle dynamics in Section \ref{sec3}.

As a special case, the electric field autocorrelation function is considered
in Section \ref{sec4} with the objective of providing a clear interpretation
for the $Z$ dependence observed in simulations. For increasing $Z$ this
dependence includes 1) an increasing covariance of the field (initial value
of the correlation function), 2) a decreasing correlation time, and 3) the
development of a strong domain of anti-correlation at intermediate times
\cite{ilya}. It is shown here that all three features can be attributed to
an increasing contribution from the bound states of the single particle
effective dynamics representing actual metastable trapped trajectories of
the $N$ particle dynamics. With this understanding of the active mechanisms,
a simple analytic and accurate model for the bound and free state
contributions is proposed and tested. The dynamics is restricted to circular
and straight line trajectories, and the electron - ion charge correlation is
represented by a nonlinear Debye distribution. Somewhat surprisingly, the
model reproduces all of the above $Z$ dependencies with remarkable accuracy.
This provides the basis for a practical representation of more general
correlation functions, such as the dynamic structure factor, and more
complex state conditions required for plasma spectroscopy in hot, dense
matter \cite{Stambulchik,wrighton}.

To illustrate the practical utility of the model, collective excitations are
explored briefly in Section \ref{sec6} using the model to evaluate the
dielectric function for this nonuniform electron distribution about the ion.
For weakly nonuniform conditions (small $Z$) the results are suggestive of a
local density approximation whereby the modes are similar to those of a
uniform electron gas, but with the density replaced by the actual local
density near the ion. However, this simple approximation fails for larger $Z$
where the bound states dominate and long wavelength plasmons are replaced by
local excitations at the circular orbit frequencies. Finally, some future
directions are discussed in the last Section.

\section{Correlation Functions and Markovian Approximation}

\label{sec2}

Consider a system of $N_{e}$ electrons of charge $-e$, an infinitely massive
positive ion of charge $Ze$ placed at the origin, and a rigid uniform
positive background for overall charge neutrality contained in a large
volume $V$. The Hamiltonian is
\begin{equation}
H=\sum_{\alpha =1}^{N_{e}}\left( \frac{1}{2}mv_{\alpha }^{2}+V_{ei}\left(
r_{\alpha }\right) +V_{eb}\left( r_{\alpha }\right) \right) +\frac{1}{2}%
\sum_{\alpha ,\gamma }^{N_{e}}V_{ee}(r_{\alpha \gamma })  \label{2.1}
\end{equation}%
where $\mathbf{r}_{\alpha }$ and $\mathbf{v}_{\alpha }$ are the position and
velocity of electron $\alpha $. The Coulomb interaction between electrons $%
\alpha $ and $\gamma $ is denoted by $V_{ee}(r_{\alpha \gamma })$ where $%
r_{\alpha \gamma }\equiv \left\vert \mathbf{r}_{\alpha }-\mathbf{r}_{\gamma
}\right\vert $. Also, $V_{ei}\left( r_{\alpha }\right) $ is the electron-ion
interaction for electron $\alpha $, and $V_{eb}\left( r_{\alpha }\right) $
is the Coulomb interaction for electron $\alpha $ with the uniform
neutralizing background. In a quantum description, $V_{ei}\left( r_{\alpha
}\right) $ is also a Coulomb interaction but in the classical case the short
range attractive divergence must be ``regularized" within a distance $\delta
$ of the order of the de Broglie wavelength \cite{filinov}. The simplest
such form is \cite{ebeling}%
\begin{equation}
V_{ei}\left( r_{\alpha }\right) =-\frac{Ze^{2}}{r_{\alpha }}\left(
1-e^{-r_{\alpha }/\delta }\right) .  \label{2.1a}
\end{equation}%
In the remainder of this presentation such a semi-classical description is
assumed. Comments on the corresponding quantum analysis are given in the
final Discussion section.

The typical response functions characterizing dynamical excitations in a
plasma are the charge density or current autocorrelation functions, which
are sums of single particle functions. More generally, the correlation
functions of this type are defined by
\begin{equation}
C_{AB}(t)=\left\langle A(t)B\right\rangle =\int d\Gamma A(\Gamma
_{t})B\left( \Gamma \right) \rho _{e}\left( \Gamma \right)  \label{2.2}
\end{equation}%
where $\Gamma =\left\{ x_{1},..,x_{N_{e}}\right\} $ is a point in the $%
6N_{e} $ dimensional phase space, and $x_{\alpha }=\mathbf{r}_{\alpha }$,$%
\mathbf{v}_{\alpha }$ denotes a point in the phase space of particle $\alpha
$. The notation $\Gamma _{t}$ denotes the evolution of the point $\Gamma $
to a time $t$ later under the dynamics generated by the Hamiltonian of (\ref%
{2.1}). The role of the central fixed ion is suppressed in this notation,
and it acts as an external potential for the electrons. The phase functions $%
A(\Gamma )$ and $B(\Gamma )$ denote some observables of interest, composed
of sums of single particle functions.
\begin{equation}
A=\sum_{\alpha =1}^{N_{e}}a(x_{\alpha }),\hspace{0.25in}B=\sum_{\alpha
=1}^{N_{e}}b(x_{\alpha }).  \label{2.3}
\end{equation}%
Finally, the average is over an equilibrium ensemble (e.g., Gibbs), $\rho
_{e}\left( \Gamma \right) $. Because of the special form (\ref{2.3}), the $N$
particle average can be reduced to a corresponding average in the single
electron subspace, by partial integration over $N_{e}-1$ electron degrees of
freedom (see Appendix \ref{appB})
\begin{equation}
C_{AB}(t)=\int dxn(r)\phi \left( v\right) a(x)\overline{b}(x,t).  \label{2.4}
\end{equation}%
Here, $n(r)$ is the equilibrium number density for electrons at a distance $%
r $ from the ion (the precise definition as a partial integral of $\rho
_{e}\left( \Gamma \right) $ is given in Appendix A), and $\phi \left(
v\right) $ is the Maxwell-Boltzmann velocity distribution. The function $%
\overline{b}(x,t)$ at $t=0$ is linearly related to the single particle phase
function $b(x)$ in (\ref{2.3})%
\begin{equation}
\overline{b}(x,0)=\overline{b}(x)=b(x)+\int dx^{\prime }n(\mathbf{r}^{\prime
})\phi \left( v^{\prime }\right) h\left( \mathbf{r},\mathbf{r}^{\prime
}\right) b(x^{\prime }).  \label{2.5}
\end{equation}%
The correlation function $h\left( \mathbf{r},\mathbf{r}^{\prime }\right) $
is related to the joint number density $n(\mathbf{r},\mathbf{r}^{\prime })$
for two electrons at $\mathbf{r}$ and $\mathbf{r}^{\prime }$ with the ion at
the origin by%
\begin{equation}
n(r)n(r^{\prime })h\left( \mathbf{r},\mathbf{r}^{\prime }\right) \equiv n(%
\mathbf{r},\mathbf{r}^{\prime })-n(r)n(r^{\prime }).  \label{2.6}
\end{equation}%
The precise definition for $n(\mathbf{r},\mathbf{r}^{\prime })$ as a partial
integral of $\rho _{e}\left( \Gamma \right) $ is given in Appendix \ref{appA}%
. The time evolution of $\overline{b}(x,t)$ in the single particle phase
space is governed by a linear equation of the form
\begin{equation}
\partial _{t}\overline{b}(x,t)+\int dx^{\prime }\mathcal{L}\left(
x,x^{\prime };t\right) \overline{b}(x^{\prime },t)=0.  \label{2.7}
\end{equation}%
All of the results up to this point are still exact.

The difficult many-body problem is encountered in the determination of $%
\mathcal{L}\left( x,x^{\prime };t\right) $. Weak coupling and perturbation
expansions are not appropriate for high $Z$ ions or conditions for strongly
coupled electrons so instead a Markovian approximation is proposed,
\begin{equation}
\mathcal{L}\left( x,x^{\prime };t\right) \rightarrow \mathcal{L}\left(
x,x^{\prime };t=0\right) \equiv \mathcal{L}\left( x,x^{\prime }\right) .
\label{2.8}
\end{equation}%
This approximation assumes that the exact generator for the initial dynamics
persists as the dominant form for later times as well. In this way the exact
initial correlations among electrons and with the ion are included. The
detailed form for $\mathcal{L}\left( x,x^{\prime }\right) $ is obtained in
Appendix \ref{appB} with the result
\begin{equation}
\mathcal{L}\left( x,x^{\prime }\right) =\left( \mathbf{v}\cdot \nabla _{%
\mathbf{r}}-m^{-1}\nabla _{\mathbf{r}}\mathcal{V}_{ie}\left( r\right) \cdot
\nabla _{\mathbf{v}}\right) \delta \left( x-x^{\prime }\right) +\mathbf{v}%
\cdot \nabla _{\mathbf{r}}\beta \mathcal{V}_{ee}\left( \mathbf{r},\mathbf{r}%
^{\prime }\right) \phi \left( v^{\prime }\right) n\left( r^{\prime }\right) ,
\label{2.9}
\end{equation}%
where $\mathcal{V}_{ie}\left( r\right) $ and $\mathcal{V}_{ee}\left( \mathbf{%
r},\mathbf{r}^{\prime }\right) $ are ``renormalized" electron - ion and
electron - electron interactions
\begin{equation}
\mathcal{V}_{ie}\left( r\right) \equiv -\beta ^{-1}\ln n\left( r\right) ,%
\hspace{0.25in}\mathcal{V}_{ee}\left( \mathbf{r},\mathbf{r}^{\prime }\right)
=-\beta ^{-1}c\left( \mathbf{r},\mathbf{r}^{\prime }\right) .  \label{2.10}
\end{equation}%
The direct correlation function $c\left( \mathbf{r},\mathbf{r}^{\prime
}\right) $ is defined in terms of $h\left( \mathbf{r},\mathbf{r}^{\prime
}\right) $ by
\begin{equation}
c\left( \mathbf{r},\mathbf{r}^{\prime }\right) =h\left( \mathbf{r},\mathbf{r}%
^{\prime }\right) -\int d\mathbf{r}^{\prime \prime }h\left( \mathbf{r},%
\mathbf{r}^{\prime \prime }\right) n\left( r^{\prime \prime }\right) c\left(
\mathbf{r}^{\prime \prime },\mathbf{r}^{\prime }\right) .  \label{2.11}
\end{equation}%
At $Z=0$ this becomes the usual Ornstein - Zernicke equation \cite{hansen}.

To interpret (\ref{2.9}), substitute this approximation into (\ref{2.7}) to
get the Markovian linear kinetic equation for $\overline{b}(x,t)$%
\begin{equation}
\left( \partial _{t}+\mathbf{v}\cdot \nabla _{\mathbf{r}}-m^{-1}\nabla _{%
\mathbf{r}}\mathcal{V}_{ie}\left( r\right) \cdot \nabla _{\mathbf{v}}\right)
\overline{b}(x,t)=- \mathbf{v}\cdot \nabla _{\mathbf{r}}\beta \int
dx^{\prime }\mathcal{V}_{ee}\left( \mathbf{r},\mathbf{r}^{\prime }\right)
\phi \left( v^{\prime }\right) n\left( r^{\prime }\right) \overline{b}%
(x^{\prime },t).  \label{2.12}
\end{equation}%
At weak electron - electron coupling $\mathcal{V}_{ee}\left( \mathbf{r},%
\mathbf{r}^{\prime }\right) \rightarrow V_{ee}(\left\vert \mathbf{r}-\mathbf{%
r}^{\prime }\right\vert )$ and at weak electron - ion coupling $\mathcal{V}%
_{ie}\left( r\right) \rightarrow V_{ie}\left( r\right) ,$ and (\ref{2.12})
is recognized as the linear Vlasov equation. More generally, the Markov
approximation (\ref{2.12}) upgrades this mean field result to include the
effects of equilibrium correlations on all interaction potentials. Thus it
is suitable for a discussion of the strong coupling conditions that occur
for $Z>1$. The left side of (\ref{2.12}) describes single electron motion
about the ion in the effective potential $\mathcal{V}_{ie}$, while the right
side describes dynamical screening of this motion.

In summary, the description of electron dynamical correlations and
fluctuations has been reduced in the Markovian approximation to
\begin{equation}
C_{AB}(t)=\int dxn(r)\phi \left( v\right) a(x)e^{-\mathcal{L}t}\overline{b}%
(x),  \label{2.13}
\end{equation}%
where $\mathcal{L}$ is the operator whose kernel is (\ref{2.9}). This
operator requires as input the equilibrium electron density $n\left(
r\right) $ and the equilibrium direct correlation function. The kinetic
equation can be solved exactly in terms of the single particle trajectories
about the ion and dielectric function for an inhomogeneous electron gas,
describing the dynamical screening due to interactions among the electrons
in the presence of the ion. The details are carried out in Appendix \ref%
{appC}, and the correlation functions are obtained from that solution in
Appendix \ref{appD}. For the class of correlation functions for which $%
a(x)=a(\mathbf{r})$ (i.e., is independent of the velocity) the Laplace
transform of (\ref{2.13}) takes the simpler form%
\begin{equation}
\int_{0}^{\infty }dte^{-zt}C_{AB}(t)=\int d\mathbf{r}d\mathbf{v}n\left(
r\right) \phi \left( v\right) a(\mathbf{r};z)\mathcal{G}_{0}(z)\overline{b}%
(x),  \label{2.14}
\end{equation}%
The dynamics is governed by the resolvent operator
\begin{equation}
\mathcal{G}_{0}\left( z\right) =\left( z+\mathcal{L}_{0}\right) ^{-1},%
\hspace{0.25in}\mathcal{L}_{0}=\mathbf{v}\cdot \nabla _{\mathbf{r}%
}-m^{-1}\nabla _{\mathbf{r}}\mathcal{V}_{ie}\left( r\right) \cdot \nabla _{%
\mathbf{v}}.  \label{2.15}
\end{equation}%
The generator for the dynamics, $\mathcal{L}_{0},$ is seen to be that for a
single electron interacting with the ion via the effective mean field
potential $\mathcal{V}_{ie}\left( r\right) $. The function $a(\mathbf{r};z)$
is the given function $a(\mathbf{r})$, modified by dynamical screening%
\begin{equation}
a(\mathbf{r};z)=\int d\mathbf{r}^{\prime }a(\mathbf{r}^{\prime })\epsilon
^{-1}\left( \mathbf{r}^{\prime },\mathbf{r};z\right) ,  \label{2.16}
\end{equation}%
where $\epsilon \left( \mathbf{r},\mathbf{r^\prime};z\right) $ is the
``dielectric function" for the electrons in the presence of the ion \cite%
{dielectric}%
\begin{equation}
\epsilon \left( \mathbf{r},\mathbf{r}^{\prime };z\right) =\delta \left(
\mathbf{r}-\mathbf{r}^{\prime }\right) -\int d\mathbf{r}^{\prime \prime }\pi
\left( \mathbf{r},\mathbf{r}^{\prime \prime };z\right) \mathcal{V}_{ee}(%
\mathbf{r}^{\prime \prime },\mathbf{r}^{\prime }),  \label{2.17}
\end{equation}%
and $\pi \left( \mathbf{r},\mathbf{r}^{\prime \prime };z\right) $ is
\begin{equation}
\pi \left( \mathbf{r},\mathbf{r}^{\prime \prime };z\right) =-\beta n(r)\int d%
\mathbf{v}\phi \left( v\right) \mathcal{G}_{0}(z)\mathbf{v}\cdot \mathbf{%
\nabla }_{\mathbf{r}}\delta \left( \mathbf{r}-\mathbf{r}^{\prime \prime
}\right) .  \label{2.17a}
\end{equation}

\subsection{Dynamic structure factor}

An important example is the autocorrelation function for the electron
density near the ion. In the absence of the ion this is referred to as the
dynamic structure factor and that terminology will be used here in the
presence of the ion as well. The correlation function $C_{AB}(t)=C(\mathbf{%
q,q}^{\prime };t)$ is constructed from the local densities of (\ref{2.3})
with $a(x_{\alpha })=\delta \left( \mathbf{q}-\mathbf{r}_{\alpha }\right) $
and $b(x_{\alpha })=\delta \left( \mathbf{q}^{\prime }-\mathbf{r}_{\alpha
}\right) $. Then (\ref{2.14}) becomes%
\begin{equation}
\int_{0}^{\infty }dte^{-zt}C(\mathbf{q,q}^{\prime };t)=\int d\mathbf{r}d%
\mathbf{v}n\left( r\right) \phi \left( v\right) \epsilon ^{-1}\left( \mathbf{%
q},\mathbf{r};z\right) \mathcal{G}_{0}\left( z\right) s\left( \mathbf{r},%
\mathbf{q}^{\prime }\right) .  \label{2.19}
\end{equation}%
Here $s\left( \mathbf{r},\mathbf{q}^{\prime }\right) $ is the static
structure factor%
\begin{eqnarray}
s\left( \mathbf{r},\mathbf{q}^{\prime }\right) &=&\delta \left( \mathbf{r}-%
\mathbf{q}^{\prime }\right) +n\left( q^{\prime }\right) h\left( \mathbf{r},%
\mathbf{q}^{\prime }\right) =\delta \left( \mathbf{r}-\mathbf{q}^{\prime
}\right) +n\left( q^{\prime }\right) h\left( \mathbf{q}^{\prime },\mathbf{r}%
\right)  \nonumber \\
&=&\epsilon ^{-1}\left( \mathbf{q}^{\prime },\mathbf{r};0\right) ,
\label{2.19a}
\end{eqnarray}%
representing the exact initial correlations. The last equality of (\ref%
{2.19a}) is proved in Appendix \ref{appD}.

Equation (\ref{2.19}) is the exact short time (Markovian) form for the
dynamic structure factor. For $Z=0$ it becomes the usual random phase
approximation (RPA) with \textquotedblleft local field corrections"; this
means that the bare electron - electron potential has been replaced by $%
\mathcal{V}_{ee}$ (the corresponding direct correlation function) to account
for exact initial correlations. The $Z=0$ case has been studied in detail
for the hydrogen plasma, where this approximation is found to be very good
up to moderate plasma coupling strengths over a wide range of space and time
scales \cite{cauble}. Equation (\ref{2.15}) extends this approximation to
include the presence of the ion for $Z\neq 0$, corresponding to an
inhomogeneous RPA.

\subsection{Electric field autocorrelation function}

A second important example is the autocorrelation function $C(t)$ for the
electron electric field at the ion, where%
\begin{equation}
A=B=\sum_{\alpha =1}^{N_{e}}\mathbf{e}\left( \mathbf{r}_{\alpha }\right) ,%
\hspace{0.4in}\mathbf{e}\left( \mathbf{r}_{\alpha }\right) =\nabla _{\mathbf{%
r}_{\alpha }}V_{ei}\left( r_{\alpha }\right) .  \label{2.20}
\end{equation}%
This correlation function also is obtained from the dynamic structure factor
$C(\mathbf{r,r}^{\prime };t)$ by integration%
\[
C(t)=\int d\mathbf{r}d\mathbf{r}^{\prime }\mathbf{e}\left( \mathbf{r}\right)
C(\mathbf{r,r}^{\prime };t)\mathbf{e}\left( \mathbf{r}^{\prime }\right) ,
\]%
so
\begin{equation}
\int_{0}^{\infty }dte^{-zt}C(t)=\int d\mathbf{r}d\mathbf{v}n\left( r\right)
\phi \left( v\right) \mathbf{e}(\mathbf{r};z)\cdot \mathcal{G}_{0}\left(
z\right) \mathbf{e}_{s}(\mathbf{r}).  \label{2.22a}
\end{equation}%
Interestingly, one of the fields is dynamically screened while the other is
statically screened,%
\begin{equation}
\mathbf{e}(\mathbf{r};z)=\int d\mathbf{r}^{\prime }\mathbf{e}(\mathbf{r}%
^{\prime })\epsilon ^{-1}\left( \mathbf{r}^{\prime },\mathbf{r};z\right) ,%
\hspace{0.25in}\mathbf{e}_{s}(\mathbf{r})=\int d\mathbf{r}^{\prime }\mathbf{e%
}(\mathbf{r}^{\prime })s\left( \mathbf{r}^{\prime },\mathbf{r}\right) =%
\mathbf{e}(\mathbf{r};z=0).  \label{2.22}
\end{equation}

\section{Bound and free contributions}

\label{sec3}

The trajectories of the mean field generator $\mathcal{L}_{0}$ do not
represent the dynamics of any given electron, but rather their effective
collective representation. Thus, the interactions among many electrons
appears in the mean field theory only through their modification of the
potentials. The bound states of the effective potential $\mathcal{V}%
_{ie}\left( r\right) $ are representations of real metastable states in the
MD simulation. At weak coupling there are few such metastable states and
their lifetimes are short compared to the correlation time for the field
autocorrelation function. As $Z$ increases the stronger coupling gives rise
to more metastable states with longer lifetimes. There is a crossover of
these lifetimes to values larger than the correlation time at which point
they behave essentially as bound states for the relevant time scales. To
isolate the effects of such bound states it is useful to divide the phase
space integral of (\ref{2.13}) into contributions from bound and free parts
of that phase space. The decomposition is defined by the negative and
positive energy states for the effective potential $\mathcal{V}_{ie}\left(
r\right) $. For a given position $r$ there is a maximum velocity $v_{m}(r)$
above which the total energy is positive%
\begin{equation}
v_{m}(r)=\sqrt{-2\mathcal{V}_{ie}\left( r\right) /m}.  \label{2.25}
\end{equation}%
The single particle equilibrium density for the position and velocity can
therefore be divided into two contributions
\begin{eqnarray}
n(r)\phi \left( v\right) &=&\Theta \left( v_{m}(r)-v\right) n(r)\phi \left(
v\right) +\Theta \left( v-v_{m}(r)\right) n(r)\phi \left( v\right)  \nonumber
\\
&=&\left( n(r)\phi \left( v\right) \right) _{b}+\left( n(r)\phi \left(
v\right) \right) _{f}.  \label{2.26}
\end{eqnarray}%
For example, integration over the velocity gives the relative contribution
of bound states to the density $n(r)$
\begin{eqnarray}
n_{b}(r) &\equiv &\int d\mathbf{v}\left( n(r)\phi \left( v\right) \right)
_{b}  \nonumber \\
&=&n(r)4\pi ^{-\frac{1}{2}}v_{0}^{-3}\int_{0}^{v_{m}(r)}dvv^{2}e^{-\left(
v/v_{0}\right) ^{2}}  \nonumber \\
&=&n(r)\left( \mathrm{erf}\left( \frac{v_{m}(r)}{v_{0}}\right) -\frac{2}{%
\sqrt{\pi }}\frac{v_{m}(r)}{v_{0}}e^{-\left( v_{m}(r)/v_{0}\right)
^{2}}\right) .  \label{2.27}
\end{eqnarray}%
where $v_{0}=\sqrt{2k_{B}T/m_{e}}$ is the thermal velocity of the electron
and $\mathrm{erf}(x)$ is the error function. Since $v_{m}(r)$ is
proportional to $\sqrt{Z}$, $n_{b}(r)$ is an increasing function of $Z$ for
all $r$.

The decomposition (\ref{2.26}) provides the identification of contributions
to the correlation functions from bound and free states
\begin{equation}
C_{AB}(t)=C_{AB}^{b}(t)+C_{AB}^{f}(t).  \label{2.29}
\end{equation}%
The analysis of the following Sections shows that the interesting $Z$
dependence of electron dynamics can be understood in terms of the relative
sizes of these two contributions.

\section{Example: Electric Field Autocorrelation Function}

\label{sec4}

The dielectric function $\epsilon \left( \mathbf{r},\mathbf{r}^{\prime
};z\right) $ describes a crossover from no screening at short times (large $%
z $) to static screening at large times ($z\rightarrow 0$). \ In the next
two sections, the short time form with $\epsilon \left( \mathbf{r},\mathbf{r}%
^{\prime };z\right) \rightarrow \delta \left( \mathbf{r}-\mathbf{r}^{\prime
}\right) $ will be considered. In that case the correlation functions in (%
\ref{2.14}) become the effective single particle functions
\begin{equation}
C_{AB}(t)\rightarrow \int d\mathbf{r}d\mathbf{v}n\left( r\right) \phi \left(
v\right) a(\mathbf{r})e^{-\mathcal{L}_{0}t}\overline{b}(x),  \label{3.1}
\end{equation}%
which are asymptotically exact at short times. The effects of dynamical
screening at longer times are discussed briefly in Section \ref{sec6}.

The electric field autocorrelation function is particularly instructive
since the field is sensitive to configurations closest to the ion. Also, its
time integral determines the dominant contribution to the half width of
spectral line widths broadened by electrons in many practical cases\cite%
{lewis}. The short time form of (\ref{2.22a}) is
\begin{equation}
C(t)\rightarrow \int d\mathbf{r}d\mathbf{v}n\left( r\right) \phi \left(
v\right) \mathbf{e}(\mathbf{r})\cdot e^{-\mathcal{L}_{0}t}\mathbf{e}_{s}(%
\mathbf{r}).  \label{3.2}
\end{equation}%
It is shown in Appendix \ref{appA} that the statically screened field of (%
\ref{2.22}) simplifies further to%
\begin{equation}
\mathbf{e}_{s}(\mathbf{r})=\frac{1}{Ze}\nabla _{\mathbf{r}}\mathcal{V}%
_{ie}\left( r\right) .  \label{3.3}
\end{equation}%
Thus, the number density $n\left( r\right) $ determines all of the
ingredients needed for calculation of $C(t)$. It is calculated here in the
HNC approximation described in Appendix \ref{appA}. The electron - electron
coupling strength is measured by the dimensionless ratio $\Gamma =\beta
e^{2}/r_{0}$ where $r_{0}$ is the average distance between electrons,
determined from the density by $4\pi n_{e}r_{0}^{3}/3=1$. The ion - electron
coupling is measured by $\sigma =-\beta V_{ei}\left( 0\right) =\beta
Ze^{2}/\delta =Z\Gamma \left( r_{0}/\delta \right) $. The results presented
here are for $\Gamma =0.1$, and $\sigma =0.25Z$ for values of $Z\leq 40$.
The corresponding quantum regularization length is $\delta /r_{0}=0.4$. The
electron - electron coupling is therefore weak, but the ion - electron
coupling can be very strong, $\sigma \leq 10$. These conditions were chosen
because previous molecular dynamics studies have been performed at these
values \cite{ilya}.

\begin{figure}[t]
\includegraphics[width=0.8\columnwidth]{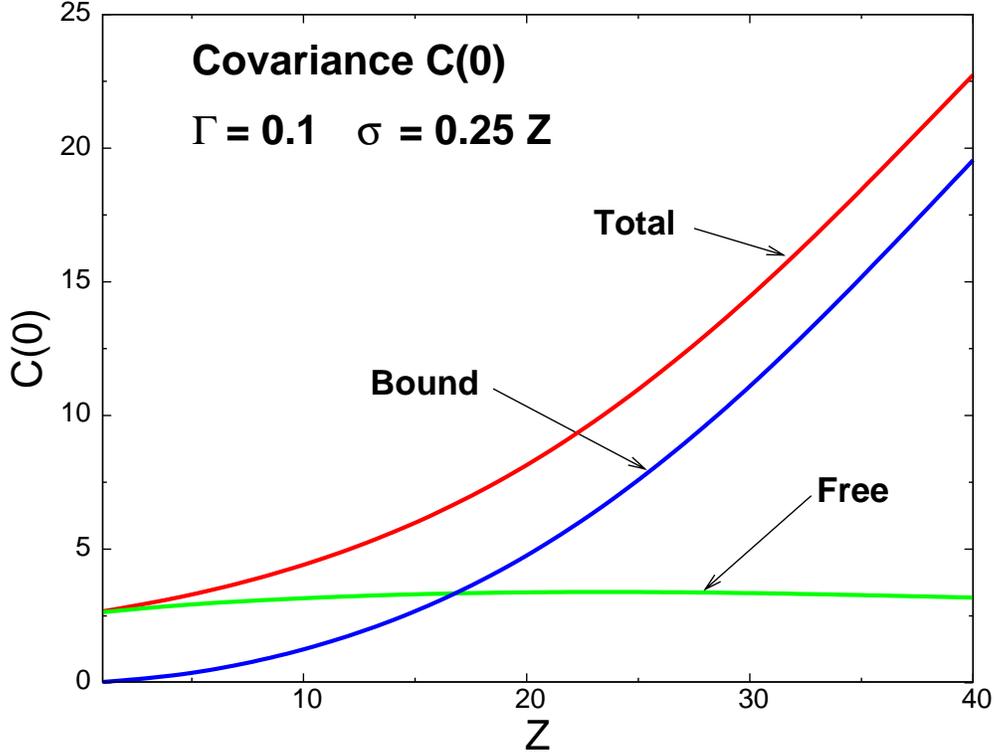} .
\caption{Bound and free state contributions to the field covariance $C(0)$
as a function of $Z$.}
\label{fig1}
\end{figure}

It is useful to anticipate the increasing role of bound states with
increasing $Z$ by considering first the time independent covariance $C(0)$.
This is shown in Figure \ref{fig1}. The sharp increase above $Z\sim 5$ is
seen to be entirely due to the appearance of the bound states. Similar
strong effects on dynamical structure are observed.
\begin{figure}[t]
\includegraphics[width=0.8\columnwidth]{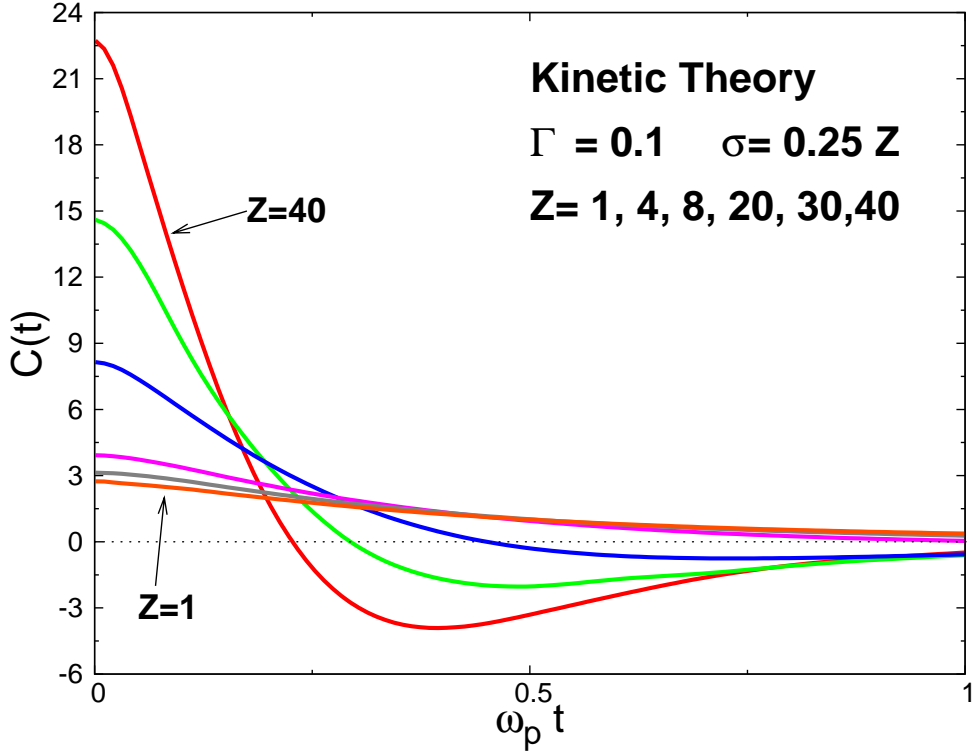} .
\caption{Autocorrelation function for an electron field at an ion of charge
number $Z=1,4,8,20,30,$ and $40$}
\label{fig2}
\end{figure}
Figure \ref{fig2} shows the results for $C(t)$ calculated from (\ref{3.2})
for $Z=1,4,8,20,30,40$. The development of a strong anti-correlation and the
decreasing initial correlation time with increasing $Z$ is evident. These
are the effects noted above, first observed in MD simulations \cite{ilya}.
The interpretation of this two-fold dependence on $Z$ is provided by Figures %
\ref{fig3} and \ref{fig4} showing the contributions from bound and free
state contributions for $Z=4,30$. For $Z=4$ the dominant contribution is
from free states, which have a monotonic positive decay. In contrast, for $%
Z=30$ the dominant contribution is from bound states which provide the
negative anti-correlation as the sign of the field changes along each
trajectory when it passes through apsidal distances. The time for this
change can be estimated by half the period for a circular orbit at position $%
r$, which is proportional to $\left( r^{3}/Z\right) ^{1/2}$. This is
consistent with the observed decrease in correlation time in Figure \ref%
{fig2}. Further elaboration and explanation is provided by the simple model
of the next section.
\begin{figure}[t]
\includegraphics[width=0.8\columnwidth]{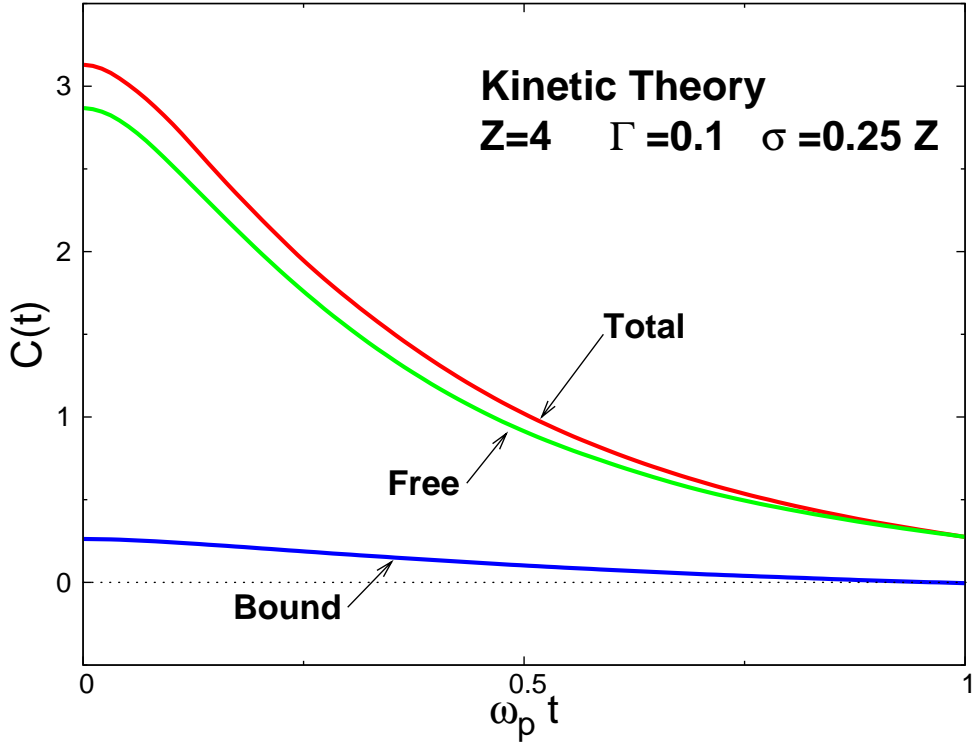} .
\caption{Bound and free state contributions to $C(t)$ for $Z=4$.}
\label{fig3}
\end{figure}
\begin{figure}[t]
\includegraphics[width=0.8\columnwidth]{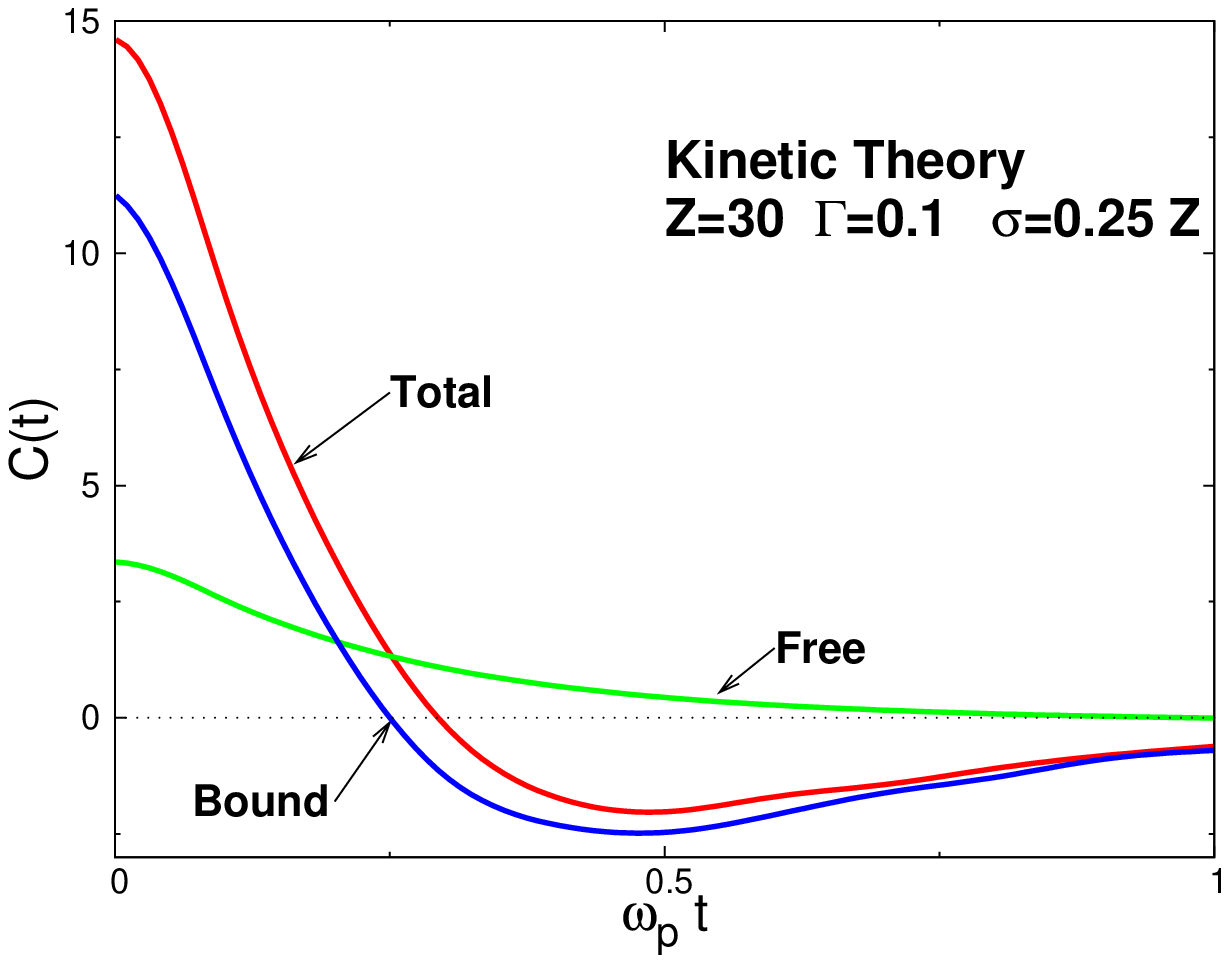} .
\caption{Bound and free state contributions to $C(t)$ for $Z=30$.}
\label{fig4}
\end{figure}
\begin{figure}[t]
\includegraphics[width=0.8\columnwidth]{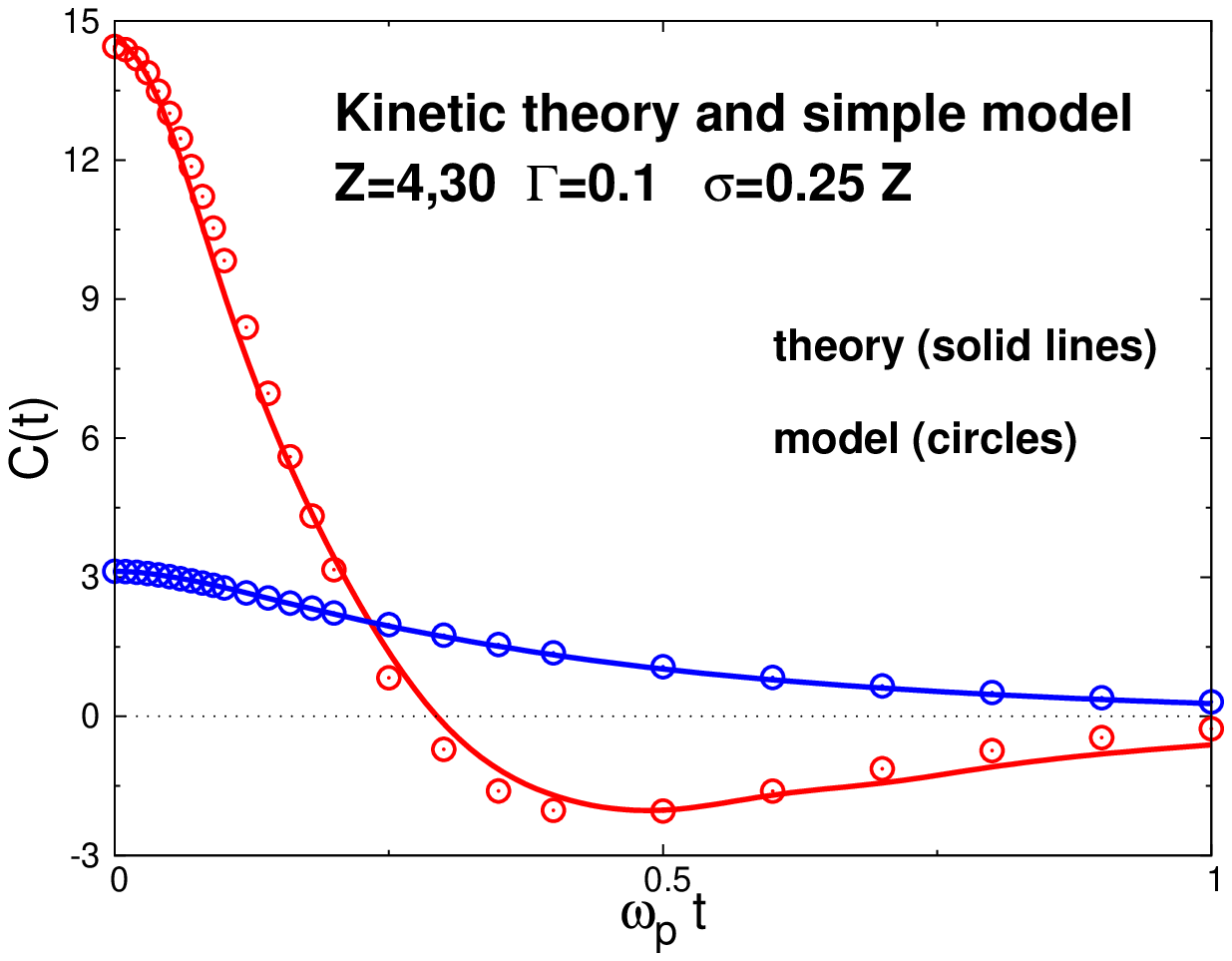} .
\caption{Comparison of $C(t)$ calculated from kinetic theory and from the
simple model for $Z=4$ and $30$.}
\label{fig5}
\end{figure}

The results of this section demonstrate the utility of the kinetic theory
for conditions of strong ion - electron coupling. Although only weak
electron - electron coupling was considered, the theory is applicable to
strong coupling among electrons as well. Also, while attention in this
section has been limited to the electric field autocorrelation function it
is clear that the analysis applies with equal ease to the dynamic structure
factor as well, with only the additional complication of more parameters
(i.e., two position vectors) characterizing that function. The decomposition
of the correlation function into bound and free contributions demonstrates
that the interesting features associated with increasing $Z$ can be
attributed entirely to the increasing contribution from bound states.

\section{A simple, analytic, and accurate model}

\label{sec5}

Consider again the electric field auto correlation function given by (\ref%
{3.2}) explicitly decomposed into its bound and free contributions%
\begin{equation}
C(t)=C^{b}(t)+C^{f}(t),  \label{4.1}
\end{equation}%
\begin{equation}
C^{b,f}(t)\rightarrow \int d\mathbf{r}d\mathbf{v}\left( n\left( r\right)
\phi \left( v\right) \right) _{b,f}\mathbf{e}(\mathbf{r})\cdot e^{-\mathcal{L%
}_{0}t}\mathbf{e}_{s}(\mathbf{r}),  \label{4.2}
\end{equation}%
The objective here is to capture the qualitative features of the bound and
free contributions in a very simple model that allows further elaboration of
their relative roles and the mechanisms involved. This is accomplished by
assuming circular trajectories for the bound states and straight line
trajectories for the free states,
\begin{equation}
C^{f}(t)\rightarrow \int d\mathbf{r}d\mathbf{v}\left( n\left( r\right) \phi
\left( v\right) \right) _{f}\mathbf{e}(\mathbf{r})\cdot \mathbf{e}_{s}(%
\mathbf{r-v}t),  \label{4.3}
\end{equation}%
\begin{equation}
C^{b}(t)\rightarrow \int d\mathbf{r}d\mathbf{v}\left( n\left( r\right) \phi
\left( v\right) \right) _{b}e(r)e_{s}(r)\cos \left( \frac{v_{c}(r)}{r}%
t\right) .  \label{4.4}
\end{equation}%
Here $\cos (v_{c}(r)t/r)=\widehat{\mathbf{r}}\cdot \widehat{\mathbf{r}}%
\left( t\right) $ and $r(t)=r$ for circular obits. In this case the velocity
must be orthogonal to $\mathbf{r}$ with the specified magnitude%
\begin{equation}
v_{c}\left( r\right) =\sqrt{\frac{r}{m}\frac{d\mathcal{V}_{ie}\left(
r\right) }{dr}},  \label{4.5}
\end{equation}%
for consistency with Newton's equations. The Maxwellian in $\left( n\left(
r\right) \phi \left( v\right) \right) _{b}$ must therefore be replaced by
this restriction on the velocities
\begin{equation}
\left( n\left( r\right) \phi \left( v\right) \right) _{b}\rightarrow \frac{1%
}{2\pi }n_{b}\left( r\right) \Theta \left( v_{m}(r)-v\right) \delta \left(
\mathbf{v\cdot }\widehat{\mathbf{r}}\right) \delta \left( u-v_{c}\left(
r\right) \right) ,  \label{4.6}
\end{equation}%
where $\mathbf{u}$ is the component of $\mathbf{v}$ orthogonal to $\mathbf{r}
$. The factor $n_{b}(r)$ is given by (\ref{2.27}) and is required by the
correct normalization on integration over all velocities. Equation (\ref{4.4}%
) then becomes
\begin{equation}
C^{b}(t)\rightarrow 4\pi \int_{0}^{\infty }drr^{2}n_{b}\left( r\right)
e(r)e_{s}(r)\cos (\omega _{c}(r)t)  \label{4.7}
\end{equation}%
where $\omega _{c}(r)\equiv v_{c}(r)/r$. Note that these modifications of
the trajectories do not affect the initial value, $C(0)$, which is still
exact.

One further simplification is made to complete the \ model. The effective
potential $\mathcal{V}_{ie}\left( r\right) $ is replaced by its weak
coupling Debye form%
\begin{equation}
\mathcal{V}_{ie}\left( r\right) \rightarrow -\frac{\overline{Z}e^{2}}{\left(
1-\left( \delta /\lambda \right) ^{2}\right) }\frac{1}{r}\left(
e^{-r/\lambda }-e^{-r/\delta }\right) ,  \label{4.8}
\end{equation}%
where $\lambda =r_{0}/\sqrt{3\Gamma }$ is the Debye length and $\delta $ is
the quantum regularization length of (\ref{2.1a}). The corresponding
electron density is now the non-linear Debye form%
\begin{equation}
n\left( r\right) =n_{e}\exp \left( \beta \frac{\overline{Z}e^{2}}{\left(
1-\left( \delta /\lambda \right) ^{2}\right) }\frac{1}{r}\left(
e^{-r/\lambda }-e^{-r/\delta }\right) \right) .  \label{4.9}
\end{equation}%
This is exact in the weak coupling limit, with $\overline{Z}\rightarrow Z$.
More generally, $\overline{Z}$ is chosen to give the correct value of $C(0)$
using (\ref{4.8}) and (\ref{4.9}) in the exact equation for $C(0)$ and
adjusting $\overline{Z}$ to fit the values obtained from HNC%
\begin{equation}
C^{\ast }(0)=\frac{r_{0}^{4}}{e^{2}}C(0)=\frac{3}{\overline{Z}\Gamma }%
\int_{0}^{\infty }dyye^{-y}\left( e^{\overline{Z}f(y)}-1\right) ,
\label{4.10}
\end{equation}%
with%
\begin{equation}
f(y)=\frac{1}{y}\frac{\Gamma }{\delta ^{\ast }\left( 1-\left( \sqrt{3\Gamma }%
\delta ^{\ast }\right) ^{2}\right) }\left( e^{-\delta ^{\ast }\sqrt{3\Gamma }%
y}-e^{-y}\right) ,\hspace{0.25in}\delta ^{\ast }=\frac{\delta }{r_{0}}.
\label{4.11}
\end{equation}%
Table 1 gives the values obtained using $C^{\ast }(0)$ from the HNC
approximation, for the case $\Gamma =0.1$ and $\sigma =0.25Z$.

\begin{center}
\bigskip
\begin{tabular}{|c|c|c|}
\hline
$Z$ & $\overline{Z}$ & $C^{\ast }(0)$ \\ \hline
$1$ & $1.03$ & $2.67$ \\ \hline
$4$ & $3.76$ & $3.13$ \\ \hline
$8$ & $7.29$ & $3.92$ \\ \hline
$20$ & $16.79$ & $8.15$ \\ \hline
$30$ & $22.87$ & $14.4$ \\ \hline
$40$ & $27.20$ & $22.7$ \\ \hline
\end{tabular}

Table 1: Effective charge number $\overline{Z}$ for the Debye form
\end{center}

Figure 5 shows again the results of Figures 3 and 4 for $Z=4$ and $30$, now
including as well the results from the simple model of this section.
Remarkably, the use of the Debye form with the straight line and circular
trajectories gives an accurate representation of the kinetic theory results.
This provides the basis for a practical tool for use in more complex
conditions, as discussed in the last section, and for other correlation
functions, as illustrated in the next section.

\section{Dielectric function}

\label{sec6}

The last two sections addressed the short time form (\ref{3.1}) of the
correlation functions for which the dielectric function behaves as $\epsilon
\left( \mathbf{r},\mathbf{r}^{\prime };z\right) \rightarrow \delta \left(
\mathbf{r}-\mathbf{r}^{\prime }\right) $. For very long times (small $z$) $%
\epsilon \left( \mathbf{r},\mathbf{r}^{\prime };z\right) $ crosses over to
represent static screening. At intermediate times there are contributions
from collective excitations. Their description is more complex than for the
uniform electron gas for which the space dependence of the dielectric
function occurs only through $\mathbf{r}-\mathbf{r}^{\prime }$\textbf{.} To
simplify the discussion here consider the weak electron - electron limit
(but possibly strong ion - electron coupling) for which $\mathcal{V}_{ee}(%
\mathbf{r},\mathbf{r}^{\prime })\rightarrow \mathcal{V}_{ee}(\left\vert
\mathbf{r}-\mathbf{r}^{\prime }\right\vert )$ and define the partial
transform%
\begin{eqnarray}
\widetilde{\epsilon }\left( \mathbf{r},\mathbf{k};z\right) &\equiv &\int d%
\mathbf{r}^{\prime }e^{i\mathbf{k\cdot }\left( \mathbf{r}^{\prime }-\mathbf{r%
}\right) }\epsilon \left( \mathbf{r},\mathbf{r}^{\prime };z\right)  \nonumber
\\
&=&1+\beta n(r)\widetilde{V}_{ee}(k)\int_{0}^{\infty }dte^{-zt}\int d\mathbf{%
v}\phi \left( v\right) i\mathbf{k}\cdot \mathbf{v}\left( -t\right) e^{i%
\mathbf{k}\cdot \left( \mathbf{r}\left( -t\right) \mathbf{-r}\right) }.
\label{5.1}
\end{eqnarray}%
Collective excitations for the non-uniform system, $z\left( \mathbf{k},%
\mathbf{r}\right) $, are defined by $\widetilde{\epsilon }\left( \mathbf{r},%
\mathbf{k};z\left( \mathbf{k},\mathbf{r}\right) \right) =0$. For zero charge
number on the ion, $Z=0$, $\widetilde{\epsilon }\left( \mathbf{r},\mathbf{k}%
;z\right) $ reduces to the familiar RPA dielectric function of the uniform
electron gas%
\begin{equation}
\widetilde{\epsilon }\left( \mathbf{r},\mathbf{k};z\right) \rightarrow
\widetilde{\epsilon }_{RPA}\left( n_{e},\mathbf{k};z\right) =1+\beta n_{e}%
\widetilde{V}_{ee}(k)\int_{0}^{\infty }dte^{-zt}\int d\mathbf{v}\phi \left(
v\right) i\mathbf{k}\cdot \mathbf{v}e^{-i\mathbf{k}\cdot \mathbf{v}t}
\label{5.3a}
\end{equation}%
identifying the excitation spectrum $z\left( \mathbf{k}\right) $.

For $Z\neq 0$ the modes, $z\left( \mathbf{k},\mathbf{r}\right) $, depend on $%
\mathbf{r}$ due to the inhomogeneity caused by the ion. As an example,
consider the solutions with very large $z$. Expanding $\widetilde{\epsilon }%
\left( \mathbf{r},\mathbf{k};z\right) $ to order $1/z^{2}$ gives
\begin{equation}
z^{2}\left( \mathbf{k},\mathbf{r}\right) =-\frac{n(r)\widetilde{V}%
_{ee}(k)k^{2}}{m}+ik_{\parallel }r\frac{n(r)\widetilde{V}_{ee}(k)}{m}\beta
m\omega _{c}^{2}\left( r\right) +..,  \label{5.4}
\end{equation}%
where $k_{\parallel }=\mathbf{k}\cdot \mathbf{r/}r$ is the component of $%
\mathbf{k}$ along $\mathbf{r}$. For small $k$ the first term goes to the
square of the local plasma frequency%
\begin{equation}
\frac{n(r)k^{2}\widetilde{V}_{ee}(k)}{m}\rightarrow \frac{4\pi n(r)e^{2}}{m}%
\equiv \omega _{p}^{2}\left( r\right) ,  \label{5.4a}
\end{equation}%
and (\ref{5.4}) simplifies to%
\begin{equation}
z^{2}\left( \mathbf{k},\mathbf{r}\right) =-\omega _{p}^{2}\left( r\right)
+i\omega _{c}^{2}\left( r\right) \frac{k_{\parallel }r}{\left( k\lambda
_{D}\left( r\right) \right) ^{2}}+..,  \label{5.4b}
\end{equation}%
Here $\lambda _{D}^{2}\left( r\right) =1/\beta m\omega _{p}^{2}\left(
r\right) $ is the corresponding local Debye length. Therefore, if $%
k_{\parallel }=0$ the system supports plasmons with frequencies defined in
terms of the local density. This is suggestive of a more general
\textquotedblleft local density approximation" \cite{Gunnarson} where the
dielectric function for the uniform electron gas is modified by replacing
the uniform density with the actual non-uniform density
\begin{equation}
\widetilde{\epsilon }\left( \mathbf{r},\mathbf{k};z\right) \rightarrow
\widetilde{\epsilon }_{RPA}\left( n_{e},\mathbf{k};z\right) \Big|%
_{n_{e}=n\left( r\right) }.  \label{5.6}
\end{equation}%
However, this is not correct in general as is evident from (\ref{5.4}) for $%
k_{\parallel }\neq 0$ and the following.

To continue with the evaluation of (\ref{5.1}) separate into bound and free
contributions
\begin{eqnarray}
\widetilde{\epsilon }\left( \mathbf{r},\mathbf{k};z\right) &=&1+\beta n(r)%
\widetilde{\mathcal{V}}_{ee}(k)\int_{0}^{\infty }dte^{-zt}\frac{d}{dt}\left[
\int d\mathbf{v}\phi \left( v\right) \Theta \left( v-v_{m}(r)\right) e^{i%
\mathbf{k}\cdot \left( \mathbf{r}\left( -t\right) \mathbf{-r}\right) }\right.
\nonumber \\
&&\left. +\frac{1}{2\pi }f_{b}(r)\int d\mathbf{v}\Theta \left(
v_{m}(r)-v\right) \delta \left( \mathbf{v\cdot }\widehat{\mathbf{r}}\right)
\delta \left( u-v_{c}\left( r\right) \right) e^{i\mathbf{k}\cdot \left(
\mathbf{r}\left( -t\right) \mathbf{-r}\right) }\right] .  \label{5.7}
\end{eqnarray}%
Next, introduce the approximate trajectories of the last section. The
analysis is straightforward but lengthy so only the final result is given
here%
\begin{equation}
\widetilde{\epsilon }\left( \mathbf{r},\mathbf{k};z\right) =1+\beta
\widetilde{\mathcal{V}}_{ee}(k)\left( n_{f}(r)I_{f}(k,r,z)+n_{b}(r)I_{b}(%
\mathbf{k},\mathbf{r,}z)\right) .  \label{5.8}
\end{equation}%
The first term of the brackets, proportional to $n_{f}(r)\equiv
n(r)-n_{b}(r) $, is the contribution from free states, while the second term
proportional to $n_{b}(r)$ is that from bound states (recall (\ref{2.27})
for the bound state contribution $n_{b}(r)$). The functions $I_{f}(k,r,z)$
and $I_{b}(\mathbf{k},\mathbf{r,}z)$ are
\begin{equation}
I_{f}(k,r,z)=kv_{0}\int_{0}^{\infty }dte^{-zt}\frac{\int_{v_{m}(r)/v_{0}}^{%
\infty }dxx^{3}e^{-x^{2}}j_{1}\left( kv_{0}tx\right) }{%
\int_{v_{m}(r)/v_{0}}^{\infty }dxx^{2}e^{-x^{2}}},  \label{5.9}
\end{equation}%
\begin{eqnarray}
I_{b}(\mathbf{k},\mathbf{r,}z) &=&\int_{0}^{\infty
}dte^{-zt}e^{ik_{\parallel }r\left( \cos (\omega _{c}(r)t)-1\right) }\Bigg[%
-ik_{\parallel }v_{c}(r)\sin (\omega _{c}(r)t)J_{0}(k_{\perp }v_{c}\left(
r\right) \sin (\omega _{c}(r)t))  \nonumber \\
&&\qquad \qquad \qquad \qquad \qquad \qquad \qquad \qquad +\ \frac{d}{dt}%
J_{0}(k_{\perp }v_{c}\left( r\right) \sin (\omega _{c}(r)t))\Bigg].
\label{5.10}
\end{eqnarray}%
The Bessel functions $j_{1}(x)$ and $J_{0}(x)$ are
\begin{equation}
j_{1}(x)=\frac{1}{x^{2}}\left( \sin x-x\cos x\right) ,\hspace{0.25in}%
J_{0}(x)=\frac{1}{2\pi }\int_{0}^{2\pi }d\phi e^{ix\cos \phi }.  \label{5.11}
\end{equation}%
The free state contribution depends only on the magnitudes $k,r$ while the
bound state contribution depends on the directions as well, where $%
k_{\parallel }$ and $k_{\perp }$ are the components of $\mathbf{k}$ parallel
and perpendicular to $\mathbf{r}$.

Consider the small $k$, long wavelength limit of this expression for $%
\widetilde{\epsilon }\left( \mathbf{r},\mathbf{k};z\right) $. Retaining the
leading order contributions to (\ref{5.9}) and (\ref{5.10}) gives%
\begin{eqnarray}
\widetilde{\epsilon }\left( \mathbf{r},\mathbf{k};z\right) &\rightarrow
&1+\allowbreak \frac{\omega _{p}^{2}\left( r\right) }{z^{2}}\left( \frac{%
n_{f}(r)}{n(r)}+\frac{4}{3\sqrt{\pi }}\left( \frac{v_{m}(r)}{v_{0}}\right)
^{3}e^{-\left( v_{m}(r)/v_{0}\right) ^{2}}\right)  \nonumber \\
&&+\frac{\omega _{c}^{2}(r)}{z^{2}+\omega _{c}^{2}(r)}\frac{ik_{\parallel }r%
}{\left( k\lambda _{D}\left( r\right) \right) ^{2}}\frac{n_{b}(r)}{n(r)}.
\label{5.12}
\end{eqnarray}%
The limiting forms for these long wavelength excitations as functions of the
ion charge number $Z$ are%
\begin{equation}
z^{2}\rightarrow \left\{
\begin{array}{c}
-\omega _{p}^{2}\left( r\right) ,\hspace{0.25in}Z<1 \\
-\omega _{c}^{2}(r)\left( 1+\frac{ik_{\parallel }r}{\left( k\lambda
_{D}\left( r\right) \right) ^{2}}\right) ,\hspace{0.25in}Z>>1%
\end{array}%
\right.  \label{5.13}
\end{equation}%
The local plasmons are recovered for small charge numbers, while for large
charge numbers the local circular frequencies $\omega _{c}^{2}(r)$ dominate.
The complex coefficient in (\ref{5.13}) implies that these excitations are
damped.

The dielectric function $\widetilde{\epsilon }\left( \mathbf{r},\mathbf{k}%
;z\right) $ is quite complex, but it is seen that the simple model of
straight line and circular orbits simplifies this considerably. Further
analysis of the collective modes for the inhomogeneous electron gas will be
given elsewhere.

\section{Discussion}

\label{sec7}

A very general description of time correlation functions for electron
properties near a positive ion has been given by the Markovian approximation
(\ref{2.14}). The time dependence has two contributions, an effective single
particle dynamics that dominates for short and intermediate times, and a
modification of that dynamics due to collective modes. The single particle
dynamics has a strong dependence on the ion charge number $Z$, which is due
to the growing dominance of bound states for large $Z$. The description is
valid for such strong coupling conditions, since the Markovian approximation
preserves the exact equilibrium electron - ion correlations in the effective
potential governing the single particle dynamics. Similarly the electron -
electron potential is renormalized by the exact equilibrium electron -
electron correlations.

The description has been illustrated here for the special case of the
electric field autocorrelation function under the same conditions as have
been studied by MD simulations \cite{ilya}. As the time scales are short,
only the effective single particle dynamics has been considered. It remains
to explore other correlation functions for which the collective modes are
expected to be more important, and to provide a detailed characterization of
those modes via the inhomogeneous electron gas dielectric function. The
promise for progress in this direction is provided by the success of a
simple model for the bound and free state dynamics.

There are several avenues for future directions based on this work:

\begin{enumerate}
\item The analysis here is based on a semi-classical description using a
"regularized" electron - ion potential. The quality of this type of
description can be benchmarked by comparison with a corresponding quantum
description of the same effective single particle dynamics.

\item A generalization of the Markovian approximation to a two component
electron - ion plasma is straightforward. In that case the interest is in
the electron dynamics in the vicinity of one of the ions. An additional
feature is the effect of the dynamics of the ions. This constitutes a
kinetic theory for the two particle (electron and ion) distribution function.

\item The spectral line shapes from charged radiators are an important
diagnostic tool in laser fusion studies. A recent formulation of this
problem including all plasma charge correlations is expressed in terms of
constrained equilibrium time correlation functions \cite{wrighton}. The
constraint arises from a specified value of the total ion electric field
during the dynamical broadening by the electrons. The simple model described
here provides the potential for practical evaluation of these constrained
time correlation functions under the demanding conditions of hot, dense
matter. The role of charge correlations in plasma spectroscopy also has been
discussed recently in reference 8.

\item A corresponding identification of the Markov limit in a fully quantum
analysis is straightforward but the resulting renormalization of the ion -
electron and electron - electron interactions by initial correlations is
more complicated. Still, the structure obtained here of single electron
dynamics in the presence of the ion modified by collective modes of the
dielectric function remains the same.

\item The attractive distortion of the electron density by the ions,
particularly for the bound states, is a type of electron confinement. The
analysis here can be applied to real traps for charged particle confinement
(e.g., dusty plasmas near an electrode, ultra-cold plasmas in a laser trap,
valence electrons in metallic clusters, electrons in quantum dots).
Significant differences include complete confinement and relaxing charge
neutrality.
\end{enumerate}

\section{Acknowledgements}

The research was supported by the NSF/DOE Partnership in Basic Plasma
Science and Engineering under the Department of Energy award
DE-FG02-07ER54946.

\appendix

\section{Equilibrium BBGKY hierarchy}

\label{appA}

Consider a point ion of charge number $Z$ in an electron gas of average
density $n_{e}$ with a positive uniform neutralizing background of density $%
n_{b}=n_{e}$. The equilibrium structure of the electrons in the presence of
the ion is given by the one and two particle distribution functions, defined
for the equilibrium ensemble by%
\begin{equation}
f_{ie}\left( r_{10},\mathbf{v}_{0},\mathbf{v}_{1}\right) \equiv N_{e}\int
dx_{2}..dx_{N_{e}}\rho _{e}\left( \Gamma \right)   \label{A.1}
\end{equation}%
\begin{equation}
f_{iee}\left( \mathbf{r}_{10},\mathbf{r}_{20},\mathbf{v}_{0},\mathbf{v}_{1},%
\mathbf{v}_{2}\right) \equiv N_{e}^{2}\int dx_{3}..dx_{N_{e}}\rho _{e}\left(
\Gamma \right)   \label{A.2}
\end{equation}%
where $\mathbf{r}_{10}=\mathbf{r}_{1}-\mathbf{r}_{0}$ is the position of an
electron at $\mathbf{r}_{1}$ relative to the ion at $\mathbf{r}_{0}$. This
position dependence reflects the fluid symmetry (rotational invariance)
about the ion. The distribution function $f_{ie}\left( r_{10},\mathbf{v}_{0},%
\mathbf{v}_{1}\right) $ obeys the equilibrium BBGKY hierarchy equation%
\begin{eqnarray}
&&\left( \mathbf{v}_{1}\cdot \mathbf{\nabla }_{1}+\mathbf{v}_{0}\cdot
\mathbf{\nabla }_{0}+m_{0}^{-1}\mathbf{F}_{ie}\left( \mathbf{r}_{10}\right)
\cdot \mathbf{\nabla }_{\mathbf{v}_{0}}+m_{e}^{-1}\mathbf{F}_{ei}\left(
\mathbf{r}_{10}\right) \cdot \mathbf{\nabla }_{\mathbf{v}_{1}}\right)
f_{ie}\left( r_{10},\mathbf{v}_{0},\mathbf{v}_{1}\right)   \nonumber \\
&=&-\int d\mathbf{r}_{2}d\mathbf{v}_{2}\left( m_{0}^{-1}\mathbf{F}%
_{ie}\left( \mathbf{r}_{20}\right) \cdot \mathbf{\nabla }_{\mathbf{v}%
_{0}}+m_{e}^{-1}\mathbf{F}_{ee}\left( \mathbf{r}_{21}\right) \cdot \mathbf{%
\nabla }_{\mathbf{v}_{1}}\right) f_{iee}\left( \mathbf{r}_{10},,\mathbf{r}%
_{20},\mathbf{v}_{0},\mathbf{v}_{1},\mathbf{v}_{2}\right)   \nonumber \\
&&+\int d\mathbf{r}_{2}\left( m_{0}^{-1}\mathbf{F}_{ie}\left( \mathbf{r}%
_{20}\right) \cdot \mathbf{\nabla }_{\mathbf{v}_{0}}+m_{e}^{-1}\mathbf{F}%
_{ee}\left( \mathbf{r}_{21}\right) \cdot \mathbf{\nabla }_{\mathbf{v}%
_{1}}\right) f_{ie}\left( r_{10},\mathbf{v}_{0},\mathbf{v}_{1}\right) n_{b}.
\label{A.4}
\end{eqnarray}%
The last term on the right side is due to the interaction of the ion and
electron with the uniform neutralizing background whose density is $%
n_{b}=n_{e}$. Also, $\mathbf{F}_{ie}\left( \mathbf{r}_{10}\right) $ is the
force of the electron on the ion, $\mathbf{F}_{ei}=-\mathbf{F}_{ie}$ is the
reaction force of the ion on the electron, and $\mathbf{F}_{ee}\left(
\mathbf{r}_{21}\right) $ is the force between electrons. These forces are
derived from corresponding potentials%
\[
\mathbf{F}_{ei}=-\mathbf{F}_{ie}=-\mathbf{\nabla }_{1}V_{ei}\left(
r_{10}\right) ,\hspace{0.25in}\mathbf{F}_{ee}\left( \mathbf{r}_{21}\right) =-%
\mathbf{\nabla }_{1}V_{ee}\left( r_{21}\right) .
\]%
The solutions to (\ref{A.4}) have the forms
\begin{equation}
f_{ie}\left( \mathbf{r}_{0},\mathbf{v}_{0},\mathbf{r}_{1},\mathbf{v}%
_{1}\right) =\phi _{i}(v_{0})\phi _{e}(v_{1})n\left( r_{10}\right)
\label{A.5}
\end{equation}%
\begin{equation}
f_{iee}\left( \mathbf{r}_{0},\mathbf{v}_{0},\mathbf{r}_{1},\mathbf{v}_{1},%
\mathbf{r}_{2},\mathbf{v}_{2}\right) =\phi _{i}(v_{0})\phi _{e}(v_{1})\phi
_{e}(v_{2})n\left( \mathbf{r}_{10},\mathbf{r}_{20}\right)   \label{A.6}
\end{equation}%
Here $\phi _{i}(v_{0})$ and $\phi _{e}(v_{1})$ are the Maxwellians for the
ion and electron%
\begin{equation}
\phi _{\alpha }\left( v\right) =\left( \frac{m_{\alpha }}{2\pi k_{B}T}%
\right) ^{3/2}e^{-\frac{m_{\alpha }v^{2}}{2k_{B}T}},  \label{A.3}
\end{equation}%
and $n(r_{10}),$ $n(\mathbf{r}_{10},\mathbf{r}_{20})$ are the one and two
particle electron number densities (relative to the ion) normalized to $N_{e}
$ and $N_{e}^{2}$, respectively. Use of these forms in (\ref{A.4}) gives
directly%
\begin{eqnarray}
0 &=&\mathbf{v}_{0}\cdot \Bigg(-\mathbf{\nabla }_{1}n\left( r_{10}\right)
-\beta n\left( r_{10}\right) \mathbf{\nabla }_{1}V_{ei}\left( r_{10}\right)
\nonumber \\
&&\qquad \qquad \qquad \qquad \qquad \qquad -\beta \int d\mathbf{r}_{2}\left[
n\left( \mathbf{r}_{10},\mathbf{r}_{20}\right) -n\left( r_{10}\right) n_{e}%
\right] \mathbf{\nabla }_{2}V_{ei}\left( r_{20}\right) \Bigg)  \nonumber \\
&&+\,\mathbf{v}_{1}\cdot \Bigg(\mathbf{\nabla }_{1}n\left( r_{10}\right)
+\beta n\left( r_{10}\right) \mathbf{\nabla }_{1}V_{ei}\left( r_{10}\right)
\nonumber \\
&&\qquad \qquad \qquad \qquad \qquad \qquad -\beta \int d\mathbf{r}_{2}\left[
n\left( \mathbf{r}_{10},\mathbf{r}_{20}\right) -n\left( r_{10}\right) n_{e}%
\right] \mathbf{\nabla }_{2}V_{ee}\left( r_{21}\right) \Bigg )  \label{A.7}
\end{eqnarray}%
Since the velocities of the ion and electron are independent, the following
two equations hold%
\begin{equation}
\mathbf{\nabla }_{1}\ln n\left( r_{10}\right) =-\beta \mathbf{\nabla }%
_{1}V_{ei}\left( r_{10}\right) -\beta n_{e}\int d\mathbf{r}_{2}\left[ \frac{%
n\left( \mathbf{r}_{10},\mathbf{r}_{20}\right) }{n\left( r_{10}\right) n_{e}}%
-1\right] \mathbf{\nabla }_{2}V_{ei}\left( r_{20}\right)   \label{A.8}
\end{equation}%
\begin{equation}
\mathbf{\nabla }_{1}\ln n\left( r_{10}\right) =-\beta \mathbf{\nabla }%
_{1}V_{ei}\left( r_{10}\right) +\beta n_{e}\int d\mathbf{r}_{2}\left[ \frac{%
n\left( \mathbf{r}_{10},\mathbf{r}_{20}\right) }{n\left( r_{10}\right) n_{e}}%
-1\right] \mathbf{\nabla }_{2}V_{ee}\left( r_{21}\right)   \label{A.9}
\end{equation}%
This provides two, seemingly independent, equations for the same electron
density around the ion $n_{ie}\left( r_{10}\right) $. Their equivalence
implies%
\begin{equation}
\int d\mathbf{r}_{2}\left( \mathbf{F}_{ie}\left( r_{20}\right) +\mathbf{F}%
_{ee}\left( r_{12}\right) \right) \left[ n_{ee}\left( \mathbf{r}_{10},%
\mathbf{r}_{20}\right) -n\left( r_{10}\right) n_{e}\right] =0  \label{A.10}
\end{equation}%
which means that the total external force on the system of two selected
particles, the ion and the one electron, is zero at equilibrium. In other
words, the distortion of $n_{iee}\left( \mathbf{r}_{10},\mathbf{r}%
_{20}\right) $ is just such as to enforce this condition. Both equations (%
\ref{A.8}) and (\ref{A.9}) are useful, as illustrated in the following two
subsections.

In the remainder of the Appendices and in the text, only the special case
of a massive ion fixed at the origin is considered.

\subsection{Screened electric field}

The electric field due to one electron at the ion is defined by

\begin{equation}
\mathbf{e}\left( \mathbf{r}\right) =\frac{1}{Ze}\nabla _{\mathbf{r}%
}V_{ei}\left( r\right) .  \label{A.11}
\end{equation}%
The electric field autocorrelation function of (\ref{2.22a}) depends on the
associated statically screened field%
\begin{equation}
\mathbf{e}_{s}(\mathbf{r})=\int d\mathbf{r}^{\prime }\mathbf{e}(\mathbf{r}%
^{\prime })s\left( \mathbf{r}^{\prime },\mathbf{r}\right) =\mathbf{e}(%
\mathbf{r})+\frac{1}{n\left( r\right) }\int d\mathbf{r}^{\prime }\mathbf{e}(%
\mathbf{r}^{\prime })\left( n\left( \mathbf{r},\mathbf{r}^{\prime }\right)
-n\left( r\right) n\left( r^{\prime }\right) \right) .  \label{A.12}
\end{equation}%
This dependence on the two electron density $n\left( \mathbf{r},\mathbf{r}%
^{\prime }\right) $ can be eliminated using (\ref{A.8}) to get
\begin{equation}
\mathbf{e}_{s}(\mathbf{r})=-\frac{1}{\beta Ze}\mathbf{\nabla }_{1}\ln
n\left( r\right) .  \label{A.13}
\end{equation}%
An effective potential is defined in terms of the density $n\left( r\right) $
in (\ref{2.10})%
\begin{equation}
\mathcal{V}_{ie}\left( r\right) \equiv -\beta ^{-1}\ln n\left( r\right) ,
\label{A.14}
\end{equation}%
so the screened electric field is given in terms of the gradient of this
effective potential%
\begin{equation}
\mathbf{e}_{s}(\mathbf{r})=\frac{1}{Ze}\mathbf{\nabla }_{1}\mathcal{V}%
_{ie}\left( r\right) .  \label{A.15}
\end{equation}

\subsection{Hypernetted chain approximation}

A simple and accurate method to determine the density is given by the
hypernetted chain (HNC) integral equations \cite{hansen}. They can be
obtained from the usual form for a three component plasma of electrons,
protons, and ions of charge number $Z$. Then the limit is taken of uniform
proton distribution and dilute concentration for the ions of charge $Z$.
Instead, it is useful here to state the result as an approximation to (\ref%
{A.9}). Consider the mean field limit of no electron - electron correlations

\begin{equation}
\int d\mathbf{r}_{2}\left[ \frac{n\left( \mathbf{r}_{1},\mathbf{r}%
_{2}\right) }{n\left( r_{1}\right) n_{e}}-1\right] \mathbf{\nabla }%
_{2}V_{ee}\left( r_{21}\right) \rightarrow \int d\mathbf{r}_{2}\left[ \frac{%
n\left( r_{1}\right) n\left( r_{2}\right) }{n\left( r_{1}\right) n_{e}}-1%
\right] \mathbf{\nabla }_{2}V_{ee}\left( r_{21}\right) .  \label{A.16}
\end{equation}%
Then equation (\ref{A.9}) simplifies to%
\begin{equation}
\mathbf{\nabla }_{1}\left[ \ln n\left( r_{1}\right) +\beta V_{ei}\left(
r_{1}\right) +\beta n_e\int d\mathbf{r}_{2}\left( \frac{n\left( r_{2}\right)
}{n_{e}}-1\right) V_{ee}\left( r_{21}\right) \right] =0,  \label{A.17}
\end{equation}%
or%
\begin{equation}
\ln \frac{n\left( r_{1}\right) }{n_{e}}=-\beta V_{ei}\left( r_{1}\right)
-\beta n_e \int d\mathbf{r}_{2}\left( \frac{n\left( r_{2}\right) }{n_{e}}%
-1\right) V_{ee}\left( r_{21}\right) .  \label{A.18}
\end{equation}%
An arbitrary constant has been used to assure the limit $n\left(
r_{1}\right) \rightarrow n_{e}$ when $V_{ei}\left( r_{1}\right) \rightarrow
0 $. Equation (\ref{A.18}) is an integral form of the Boltzmann - Poisson
equation.

The HNC approximation is similar, but retains electron - electron
correlations in the absence of the ion
\begin{equation}
-\beta \int d\mathbf{r}_{2}\left[ \frac{n\left( \mathbf{r}_{1},\mathbf{r}%
_{2}\right) }{n\left( r_{1}\right) n_{e}}-1\right] \mathbf{\nabla }%
_{2}V_{ee}\left( r_{21}\right) \rightarrow \int d\mathbf{r}_{2}\left[ \frac{%
n\left( r_{1}\right) n\left( r_{2}\right) }{n\left( r_{1}\right) n_{e}}-1%
\right] \mathbf{\nabla }_{2}c_{ee}\left( r_{21}\right) .  \label{A.19}
\end{equation}%
The function $c_{ee}\left( r_{21}\right) $ is the electron direct
correlation function defined in terms of the electron - electron pair
correlation function (without the ion), $n_{ee}\left( r_{21}\right) $, by
the Ornstein - Zernicke equation%
\begin{equation}
c_{ee}\left( r\right) =h_{ee}\left( r\right) -n_{e}\int d\mathbf{r}^{\prime
}h_{ee}\left( r^{\prime }\right) c_{ee}\left( \left\vert \mathbf{r}-\mathbf{r%
}^{\prime }\right\vert \right) ,\hspace{0.25in}h_{ee}\left( r\right) =\frac{%
n_{ee}\left( r\right) }{n_{e}^{2}}-1  \label{A.20}
\end{equation}%
The approximation (\ref{A.19}) in (\ref{A.9}) gives the HNC approximation
for $n\left( r\right) $
\begin{equation}
\ln \frac{n\left( r_{1}\right) }{n_{e}}=-\beta V_{ei}\left( r_{1}\right)
+n_{e}\int d\mathbf{r}_{2}\left( \frac{n\left( r_{2}\right) }{n_{e}}%
-1\right) c_{ee}\left( r_{21}\right) .  \label{A.21}
\end{equation}%
This is the same as (\ref{A.18}) except that $V_{ee}\left( r_{21}\right) $
has been replaced by $-\beta ^{-1}c_{ee}\left( r_{21}\right) $.

The electron - electron direct correlation function is determined
independently from the Ornstein - Zernicke equation (\ref{A.20}) and the HNC
approximation%
\begin{equation}
\ln \frac{n_{ee}\left( r\right) }{n_{e}^{2}}=-\beta V_{ee}\left(
r_{1}\right) +h_{ee}\left( r\right) -c_{ee}\left( r\right) .  \label{A.22}
\end{equation}%
Equations (\ref{A.20}) - (\ref{A.22}) are the HNC equations used for the
numerical calculations presented here.

\section{Evaluation of $\mathcal{L}(x,x^{\prime };t=0)$}

\label{appB}

The dynamics of $C_{AB}(t)$ is conveniently expressed in terms of the
fundamental correlation function $G(x,x^{\prime };t)$%
\begin{equation}
C_{AB}(t)=\int dxdx^{\prime }a(x)G(x,x^{\prime };t)b(x^{\prime }),
\label{a.6}
\end{equation}%
\begin{equation}
G(x,x^{\prime };t)=\left\langle f\left( x,t\right) \left( f\left( x^{\prime
}\right) -\left\langle f\left( x^{\prime }\right) \right\rangle \right)
\right\rangle ,\hspace{0.3in}f\left( x\right) =\sum_{\alpha
=1}^{N_{e}}\delta \left( x-x_{\alpha }\right) .  \label{a.7}
\end{equation}%
The initial value $G(x,x^{\prime };0)$ is easily calculated, with the result
\begin{equation}
G(x,x^{\prime };0)=n\left( r\right) \phi \left( v\right) \left( \delta
\left( x-x^{\prime }\right) +\phi \left( v^{\prime }\right) n\left(
r^{\prime }\right) h\left( \mathbf{r},\mathbf{r}^{\prime }\right) \right) .
\label{a.8}
\end{equation}%
In the last equality the two electron correlation function $h\left( \mathbf{r%
},\mathbf{r}^{\prime }\right) $ has been identified from (\ref{2.6}).
Comparison with (\ref{2.5}) shows that $\overline{b}(x)$ can be written%
\begin{equation}
\overline{b}(x)=\frac{1}{n(r)\phi (v)}\int dxG(x,x^{\prime };0)b\left(
x^{\prime }\right) .  \label{a.9}
\end{equation}%
This leads to the representation (\ref{2.4}) for $C_{AB}(t)$%
\begin{equation}
C_{AB}(t)=\int dxn(r)\phi \left( v\right) a(x)\overline{b}(x,t),
\label{a.10}
\end{equation}%
with%
\begin{equation}
\overline{b}(x,t)=\int dx^{\prime }U\left( x,x^{\prime };t\right) \overline{b%
}(x^{\prime }),  \label{a.11}
\end{equation}%
\begin{equation}
U\left( x,x^{\prime };t\right) =\frac{1}{n(r)\phi (v)}\int dx^{\prime \prime
}G(x,x^{\prime \prime };t)G^{-1}(x^{\prime \prime },x^{\prime
};0)n(r^{\prime })\phi (v^{\prime }).  \label{a.12}
\end{equation}%
The inverse of $G(x,x^{\prime };0)$ has been introduced by the definition%
\begin{equation}
\delta \left( x-x^{\prime }\right) =\int dx^{\prime \prime }G(x,x^{\prime
\prime };0)G^{-1}(x^{\prime \prime },x^{\prime };0.)  \label{a.13}
\end{equation}%
It is verified that%
\begin{equation}
G^{-1}(x,x^{\prime };0)=\frac{1}{n\left( r\right) \phi \left( v\right) }%
\delta \left( x-x^{\prime }\right) -c\left( \mathbf{r},\mathbf{r}^{\prime
}\right) ,  \label{a.14}
\end{equation}%
if $c\left( \mathbf{r},\mathbf{r}^{\prime }\right) $ obeys the equation%
\begin{equation}
c\left( \mathbf{r},\mathbf{r}^{\prime }\right) =h\left( \mathbf{r},\mathbf{r}%
^{\prime }\right) -\int d\mathbf{r}^{\prime \prime }h\left( \mathbf{r},%
\mathbf{r}^{\prime \prime }\right) n\left( \mathbf{r}^{\prime \prime
}\right) c\left( \mathbf{r}^{\prime \prime },\mathbf{r}^{\prime }\right)
\label{a.15}
\end{equation}%
which is a generalization of the Ornstein-Zernicke equation \cite{hansen}.

The formal equation for $\overline{b}(x,t)$, (\ref{2.7}), follows from
differentiation of (\ref{a.11}) with respect to time and the identification%
\begin{equation}
\mathcal{L}(x,x^{\prime };t)=-\int dx^{\prime \prime }\left( \partial
_{t}U(x,x^{\prime \prime };t)\right) U^{-1}(x^{\prime \prime },x^{\prime
};t).  \label{a.16}
\end{equation}%
This provides the desired result for identifying the Markovian approximation%
\begin{equation}
\mathcal{L}\left( x,x^{\prime }\right) \equiv \mathcal{L}\left( x,x^{\prime
};t=0\right) =-\partial _{t}U(x,x^{\prime };0),  \label{a.17}
\end{equation}%
where the property $U\left( x,x^{\prime };t=0\right) =\delta \left(
x-x^{\prime }\right) $ has been used. Equation (\ref{a.12}) gives finally
\begin{equation}
\mathcal{L}(x,x^{\prime })=-\,\frac{1}{n(r)\phi (v)}\int dx^{\prime \prime
}\partial _{t}G(x,x^{\prime \prime };t)\Big|_{t=0}G^{-1}(x^{\prime \prime
},x^{\prime };0)n(r^{\prime })\phi (v^{\prime }).  \label{a.18}
\end{equation}

It only remains to calculate the initial derivative $\partial
_{t}G(x,x^{\prime \prime };t)\mid _{t=0}$ to determine $\mathcal{L}%
(x,x^{\prime })$. To simplify the notation it is useful to denote the force
on the electron due to both the ion and the uniform positive background by $%
\mathbf{F}_{0}\left( \mathbf{r}\right) $
\begin{eqnarray}
\partial _{t}G(x,x^{\prime };t)\Big|_{t=0} &=&-\int dx_{1}\delta \left(
x_{1}-x\right) \left( \mathbf{v}_{1}\cdot \nabla _{\mathbf{r}_{1}}+m^{-1}%
\mathbf{F}_{0}\left( \mathbf{r}_{1}\right) \cdot \nabla _{\mathbf{v}%
_{1}}\right) \Bigg[\delta \left( x_{1}-x^{\prime }\right) n\left(
r_{1}\right) \phi \left( v_{1}\right)   \nonumber \\
&&\qquad \qquad +\int dx_{2}\delta \left( x_{2}-x^{\prime }\right) \phi
\left( v_{1}\right) \phi \left( v_{2}\right) \left( n\left( \mathbf{r}_{1},%
\mathbf{r}_{2}\right) -n\left( r_{1}\right) n\left( r_{2}\right) \right) %
\Bigg]  \nonumber \\
&&-\int dx_{1}dx_{2}\delta \left( x_{1}-x\right) m^{-1}\mathbf{F}_{ee}\left(
\mathbf{r}_{12}\right) \cdot \nabla _{\mathbf{v}_{1}}  \nonumber \\
&&\quad \times \Bigg[\left( \delta \left( x^{\prime }-x_{1}\right) +\delta
\left( x^{\prime }-x_{2}\right) \right) \phi \left( v_{1}\right) \phi \left(
v_{2}\right) n\left( \mathbf{r}_{1},\mathbf{r}_{2}\right)   \nonumber \\
&&\quad +\int dx_{3}\delta \left( x^{\prime }-x_{3}\right) \phi \left(
v_{1}\right) \phi \left( v_{2}\right) \phi \left( v_{3}\right) \left(
n\left( \mathbf{r}_{1},\mathbf{r}_{2},\mathbf{r}_{3}\right) -n\left( \mathbf{%
r}_{1},\mathbf{r}_{2}\right) n\left( r^{\prime }\right) \right) \Bigg]
\nonumber \\
&=&-\left( \mathbf{v}\cdot \nabla _{\mathbf{r}}+m^{-1}\mathbf{F}_{0}\left(
\mathbf{r}\right) \cdot \nabla _{\mathbf{v}}\right) G(x,x^{\prime };0)
\nonumber \\
&&\ -\nabla _{\mathbf{v}}\cdot \delta \left( x^{\prime }-x\right) \phi
\left( v\right) m^{-1}\int d\mathbf{r}_{2}\mathbf{F}_{ee}\left( \left\vert
\mathbf{r}-\mathbf{r}_{2}\right\vert \right) n\left( \mathbf{r},\mathbf{r}%
_{2}\right)   \nonumber \\
&&\quad +\beta \phi \left( v\right) \phi \left( v^{\prime }\right) \mathbf{v}%
\cdot \Bigg[\mathbf{F}_{ee}\left( \left\vert \mathbf{r}-\mathbf{r}^{\prime
}\right\vert \right) n\left( \mathbf{r},\mathbf{r}^{\prime }\right)
\nonumber \\
&&\qquad \qquad \quad +\int d\mathbf{r}_{2}\mathbf{F}_{ee}\left( \left\vert
\mathbf{r}-\mathbf{r}_{2}\right\vert \right) \left( n\left( \mathbf{r}_{1},%
\mathbf{r}_{2},\mathbf{r}^{\prime }\right) -n\left( \mathbf{r}_{1},\mathbf{r}%
_{2}\right) n\left( r^{\prime }\right) \right) \Bigg]  \label{a.19}
\end{eqnarray}%
The two integrals on the right can be performed using the hierarchy equation
(\ref{A.9}) for $n\left( r\right) $ and the corresponding next order
hierarchy equation for $n\left( \mathbf{r}_{1},\mathbf{r}_{2}\right) $. In
the current notation these are
\begin{equation}
\int d\mathbf{r}_{2}\mathbf{F}_{ee}\left( \left\vert \mathbf{r}-\mathbf{r}%
_{2}\right\vert \right) n\left( \mathbf{r},\mathbf{r}_{2}\right) =\beta ^{-1}%
\mathbf{\nabla }_{1}n\left( r_{10}\right) -n\left( r_{10}\right) \mathbf{F}%
_{0}\left( \mathbf{r}\right) ,  \label{a.19a}
\end{equation}%
\begin{equation}
\int d\mathbf{r}_{2}\mathbf{F}_{ee}\left( \left\vert \mathbf{r}-\mathbf{r}%
_{2}\right\vert \right) n\left( \mathbf{r}_{1},\mathbf{r}_{2},\mathbf{r}%
^{\prime }\right) =\beta ^{-1}\mathbf{\nabla }_{1}n\left( \mathbf{r}_{1},%
\mathbf{r}^{\prime }\right) -\left( \mathbf{F}_{0}\left( \mathbf{r}%
_{1}\right) +\mathbf{F}_{ee}\left( \mathbf{r}_{1}-\mathbf{r}^{\prime
}\right) \right) n\left( \mathbf{r}_{1},\mathbf{r}^{\prime }\right) .
\label{a.19b}
\end{equation}%
Then (\ref{A.19}) becomes
\begin{eqnarray}
\partial _{t}G(x,x^{\prime };t)\Big|_{t=0} &=&-\left( \mathbf{v}\cdot \nabla
_{\mathbf{r}}+m^{-1}\mathbf{F}_{0}\left( r\right) \cdot \nabla _{\mathbf{v}%
}\right) G(x,x^{\prime };0)  \nonumber \\
&&-\left( m^{-1}\left( \beta ^{-1}\nabla _{\mathbf{q}}\ln n\left( r\right)
\right) -m^{-1}\mathbf{F}_{0}\left( r\right) \right) \cdot \nabla _{\mathbf{v%
}}\delta \left( x^{\prime }-x\right) n\left( r\right) \phi \left( v\right)
\nonumber \\
&&+\beta \phi \left( v\right) \phi \left( v^{\prime }\right) \left( n\left(
\mathbf{r}_{1},\mathbf{r}_{2}\right) -n\left( r_{1}\right) n\left(
r_{2}\right) \right) \mathbf{v}\cdot   \nonumber \\
&&\qquad \times \left( \beta ^{-1}\nabla _{\mathbf{r}}\ln \left( n\left(
\mathbf{r}_{1},\mathbf{r}_{2}\right) -n\left( {r}_{1}\right) n\left( {r}%
_{2}\right) \right) -\mathbf{F}_{0}\left( r\right) \right) .  \label{a.20}
\end{eqnarray}%
Next, eliminate the delta function using (\ref{a.8})
\begin{eqnarray}
\partial _{t}G(x,x^{\prime };t)\Big|_{t=0} &=&-\left( \mathbf{v}\cdot \nabla
_{\mathbf{r}}+m^{-1}\mathbf{F}_{0}\left( r\right) \cdot \nabla _{\mathbf{v}%
}\right) G(x,x^{\prime };0)  \nonumber \\
&&-\left( m^{-1}\left( \beta ^{-1}\nabla _{\mathbf{r}}\ln n\left( r\right)
\right) -m^{-1}\mathbf{F}_{0}\left( r\right) \right) \cdot \nabla _{\mathbf{v%
}}G(x,x^{\prime };0)  \nonumber \\
&&+\left( m^{-1}\left( \beta ^{-1}\nabla _{\mathbf{r}}\ln n\left( r\right)
\right) -m^{-1}\mathbf{F}_{0}\left( r\right) \right)   \nonumber \\
&&\qquad \qquad \cdot \nabla _{\mathbf{v}}\phi \left( v\right) \phi \left(
v^{\prime }\right) \left( n\left( \mathbf{r},\mathbf{r}^{\prime }\right)
-n\left( r\right) n\left( r^{\prime }\right) \right)   \nonumber \\
&&+\beta \phi \left( v\right) \phi \left( v^{\prime }\right) \left( n\left(
\mathbf{r}_{1},\mathbf{r}_{2}\right) -n\left( r_{1}\right) n\left(
r_{2}\right) \right) \mathbf{v}\cdot   \nonumber \\
&&\qquad \qquad \times \left( \beta ^{-1}\nabla _{\mathbf{r}}\ln \left(
n\left( \mathbf{r}_{1},\mathbf{r}_{2}\right) -n\left( r_{1}\right) n\left(
r_{2}\right) \right) -\mathbf{F}_{0}\left( r\right) \right)   \nonumber \\
&=&-\left( \mathbf{v}\cdot \nabla _{\mathbf{r}}+m^{-1}\left( \beta
^{-1}\nabla _{\mathbf{r}}\ln n\left( r\right) \right) \cdot \nabla _{\mathbf{%
v}}\right) G(x,x^{\prime };0)  \nonumber \\
&&\qquad +\phi \left( v\right) \phi \left( v^{\prime }\right) n\left(
r\right) n\left( r^{\prime }\right) \mathbf{v}\cdot \nabla _{\mathbf{r}%
}h\left( \mathbf{r},\mathbf{r}^{\prime }\right) .  \label{a.21}
\end{eqnarray}%
Substitution of this result into (\ref{a.18}) gives the generator for
initial dynamics
\begin{eqnarray}
\mathcal{L}(x,x^{\prime }) &=&\left( \mathbf{v}\cdot \nabla _{\mathbf{r}%
}+m^{-1}\left( \beta ^{-1}\nabla _{\mathbf{r}}\ln n\left( \mathbf{r}\right)
\right) \cdot \nabla _{\mathbf{v}}\right) \delta \left( x-x^{\prime }\right)
\nonumber \\
&&-\mathbf{v}\cdot \nabla _{\mathbf{r}}\int dx^{\prime \prime }\phi \left(
v^{\prime \prime }\right) n\left( r^{\prime \prime }\right) h\left( \mathbf{r%
},\mathbf{r}^{\prime \prime }\right) G^{-1}(x^{\prime \prime },x^{\prime
};0)\phi \left( v^{\prime }\right) n\left( r^{\prime }\right)   \nonumber \\
&=&\left( \mathbf{v}\cdot \nabla _{\mathbf{r}}+m^{-1}\left( \beta
^{-1}\nabla _{\mathbf{r}}\ln n\left( r\right) \right) \cdot \nabla _{\mathbf{%
v}}\right) \delta \left( x-x^{\prime }\right)   \nonumber \\
&&-\mathbf{v}\cdot \nabla _{\mathbf{r}}\left[ h\left( \mathbf{r},\mathbf{r}%
^{\prime }\right) -\int d\mathbf{r}^{\prime \prime }n\left( r^{\prime \prime
}\right) h\left( \mathbf{r},\mathbf{r}^{\prime \prime }\right) c\left(
\mathbf{r}^{\prime \prime },\mathbf{r}^{\prime }\right) \right] \phi \left(
v^{\prime }\right) n\left( r^{\prime }\right) .  \label{a.22}
\end{eqnarray}%
Finally, using (\ref{a.15}) the result (\ref{2.9}) is obtained%
\begin{equation}
\mathcal{L}(x,x^{\prime })=\left( \mathbf{v}\cdot \nabla _{\mathbf{r}%
}+m^{-1}\left( \beta ^{-1}\nabla _{\mathbf{r}}\ln n\left( r\right) \right)
\cdot \nabla _{\mathbf{v}}\right) \delta \left( x-x^{\prime }\right) -%
\mathbf{v}\cdot \nabla _{\mathbf{r}}c\left( \mathbf{r},\mathbf{r}^{\prime
}\right) \phi \left( v^{\prime }\right) n\left( r^{\prime }\right) .
\label{a.23}
\end{equation}

\section{Solution to kinetic equation}

\label{appC}

The solution to (\ref{2.12}) can be obtained in terms of an effective single
electron dynamics by direct integration
\begin{equation}
\overline{b}(x,t)=e^{-\mathcal{L}_{0}t}\overline{b}(x)-\int_{0}^{t}d\tau e^{-%
\mathcal{L}_{0}\left( t-\tau \right) }\mathbf{v}\cdot \mathbf{\nabla }_{%
\mathbf{r}}\int d\mathbf{r}^{\prime }\beta \mathcal{V}_{ee}\left( \mathbf{r},%
\mathbf{r}^{\prime }\right) I(\mathbf{r}^{\prime },\tau ),  \label{b.1}
\end{equation}%
where the generator for the effective single particle dynamics is%
\begin{equation}
\mathcal{L}_{0}\equiv \mathbf{v}\cdot \nabla _{\mathbf{r}}-m^{-1}\nabla _{%
\mathbf{r}}\mathcal{V}_{ie}\left( r\right) \cdot \nabla _{\mathbf{v}},
\label{b.2}
\end{equation}%
and the source term $I(\mathbf{r},t)$ is
\begin{equation}
I(\mathbf{r},t)\equiv \int d\mathbf{v}\phi \left( v\right) n\left( r\right)
\overline{b}(x,t).  \label{b.3}
\end{equation}%
The initial condition $\overline{b}(x)$ is given by (\ref{2.5}). It is
convenient at this point to introduce the corresponding Laplace transform%
\begin{equation}
\widetilde{b}(x,z)=\int_{0}^{\infty }dte^{-zt}\overline{b}(x,t).  \label{b.4}
\end{equation}%
Then Laplace transformation of (\ref{b.1}) gives the equation $\widetilde{b}%
(x,z)$
\begin{equation}
\widetilde{b}(x,z)=\mathcal{G}_{0}\overline{b}(x)-\mathcal{G}_{0}\mathbf{v}%
\cdot \mathbf{\nabla }_{\mathbf{r}}\int d\mathbf{r}^{\prime }\beta \mathcal{V%
}_{ee}\left( \mathbf{r},\mathbf{r}^{\prime }\right) \widetilde{I}(\mathbf{r}%
^{\prime },z),  \label{b.5}
\end{equation}%
\begin{equation}
\widetilde{I}(\mathbf{r},z)=\int d\mathbf{v}\phi \left( v\right) n\left(
r\right) \widetilde{b}(x,z),\hspace{0.2in}\mathcal{G}_{0}=\left( z+\mathcal{L%
}_{0}\right) ^{-1}  \label{b.6}
\end{equation}

An equation for $\widetilde{I}(\mathbf{r},z)$ follows from substitution of (%
\ref{b.5}) into (\ref{b.6})%
\begin{eqnarray}
\widetilde{I}(\mathbf{r},z) &=&\int d\mathbf{v}\phi \left( v\right) n\left(
r\right) \mathcal{G}_{0}\overline{b}(x,0)-\int d\mathbf{v}\phi \left(
v\right) n(r)\mathcal{G}_{0}\mathbf{v}\cdot \mathbf{\nabla }_{\mathbf{r}%
}\int d\mathbf{r}^{\prime }\beta \mathcal{V}_{ee}\left( \mathbf{r},\mathbf{r}%
^{\prime }\right) \widetilde{I}(\mathbf{r}^{\prime },z)  \nonumber \\
&=&\widetilde{I}_{0}(\mathbf{r},z)+\int d\mathbf{r}^{\prime \prime }\pi
\left( \mathbf{r},\mathbf{r}^{\prime \prime };z\right) \int d\mathbf{r}%
^{\prime }\mathcal{V}_{ee}\left( \mathbf{r}^{\prime \prime },\mathbf{r}%
^{\prime }\right) \widetilde{I}(\mathbf{r}^{\prime },z)  \label{b.7}
\end{eqnarray}%
where $\pi \left( \mathbf{r},\mathbf{r}^{\prime \prime };z\right) $ is
\begin{equation}
\pi \left( \mathbf{r},\mathbf{r}^{\prime \prime };z\right) \equiv -\beta
n(r)\int d\mathbf{v}\phi \left( v\right) \mathcal{G}_{0}\mathbf{v}\cdot
\mathbf{\nabla }_{\mathbf{r}}\delta \left( \mathbf{r}-\mathbf{r}^{\prime
\prime }\right) ,  \label{b.8}
\end{equation}%
and
\begin{equation}
\widetilde{I}_{0}(\mathbf{r},z)\equiv n\left( r\right) \int d\mathbf{v}\phi
\left( v\right) \mathcal{G}_{0}\overline{b}(x).  \label{b.9}
\end{equation}%
This is an integral equation for $I(\mathbf{r},t)$ which can be written
\begin{equation}
\int d\mathbf{r}^{\prime }\epsilon \left( \mathbf{r},\mathbf{r}^{\prime
};z\right) \widetilde{I}(\mathbf{r}^{\prime },z)=\widetilde{I}_{0}(\mathbf{r}%
,z)  \label{b.10}
\end{equation}%
The dielectric function $\epsilon \left( \mathbf{r},\mathbf{r}^{\prime
};z\right) $ is defined by
\begin{equation}
\epsilon \left( \mathbf{r},\mathbf{r}^{\prime };z\right) =\delta \left(
\mathbf{r}-\mathbf{r}^{\prime }\right) -\int d\mathbf{r}^{\prime \prime }\pi
\left( \mathbf{r},\mathbf{r}^{\prime \prime };z\right) \mathcal{V}_{ee}(%
\mathbf{r}^{\prime \prime },\mathbf{r}^{\prime }).  \label{b.11}
\end{equation}

With these results (\ref{b.5}) becomes%
\begin{equation}
\widetilde{b}(x,z)=\mathcal{G}_{0}\left[ \overline{b}(x)-\mathbf{v}\cdot
\mathbf{\nabla }_{\mathbf{r}}\int d\mathbf{r}^{\prime \prime }\beta \mathcal{%
V}_{ee}\left( \mathbf{r},\mathbf{r}^{\prime \prime }\right) \int d\mathbf{r}%
^{\prime }\epsilon ^{-1}\left( \mathbf{r}^{\prime \prime },\mathbf{r}%
^{\prime };z\right) \widetilde{I}_{0}(\mathbf{r}^{\prime },z)\right] .
\label{b.12}
\end{equation}%
The inverse dielectric function is defined by%
\begin{equation}
\int d\mathbf{r}^{\prime \prime }\epsilon \left( \mathbf{r},\mathbf{r}%
^{\prime \prime };z\right) \epsilon ^{-1}\left( \mathbf{r}^{\prime \prime },%
\mathbf{r}^{\prime };z\right) =\delta \left( \mathbf{r}-\mathbf{r}^{\prime
}\right) =\int d\mathbf{r}^{\prime \prime }\epsilon ^{-1}\left( \mathbf{r},%
\mathbf{r}^{\prime \prime };z\right) \epsilon \left( \mathbf{r}^{\prime
\prime },\mathbf{r}^{\prime };z\right) .  \label{b.13}
\end{equation}%
Equation (\ref{b.12}) is the desired solution to the kinetic equation, in
terms of the single particle dynamics of $\mathcal{G}_{0}$, since all terms
on the right side are now explicit.

The low frequency limit $\left( z=0\right) $ of $\epsilon \left( \mathbf{r},%
\mathbf{r}^{\prime };z\right) $ has a simple form in terms of the electron
correlations. First write $\epsilon \left( \mathbf{r},\mathbf{r}^{\prime
};z\right) $ as

\begin{eqnarray}
\epsilon \left( \mathbf{r},\mathbf{r}^{\prime };z\right) &=&\delta \left(
\mathbf{r}-\mathbf{r}^{\prime }\right) -\int d\mathbf{r}^{\prime \prime }\pi
\left( \mathbf{r},\mathbf{r}^{\prime \prime };z\right) \mathcal{V}_{ee}(%
\mathbf{r}^{\prime \prime },\mathbf{r}^{\prime })  \nonumber \\
&=&\delta \left( \mathbf{r}-\mathbf{r}^{\prime }\right) +\beta n(r)\int d%
\mathbf{v}\phi \left( v\right) \mathcal{G}_{0}\left( \mathcal{G}%
_{0}^{-1}-z\right) \mathcal{V}_{ee}(\mathbf{r},\mathbf{r}^{\prime })
\nonumber \\
&=&\delta \left( \mathbf{r}-\mathbf{r}^{\prime }\right) +\beta n(r)\left(
\mathcal{V}_{ee}(\mathbf{r},\mathbf{r}^{\prime })-z\int d\mathbf{v}\phi
\left( v\right) \mathcal{G}_{0}\mathcal{V}_{ee}(\mathbf{r},\mathbf{r}%
^{\prime })\right) .  \label{b.14}
\end{eqnarray}%
where the definition of $\pi \left( \mathbf{r},\mathbf{r}^{\prime };z\right)
$ in (\ref{b.8}) and $\mathcal{G}_{0}$ in (\ref{b.6}) have been used. Then
taking the real and imaginary parts of $z$ going to zero gives%
\begin{equation}
\epsilon \left( \mathbf{r},\mathbf{r}^{\prime };z=0\right) =\delta \left(
\mathbf{r}-\mathbf{r}^{\prime }\right) +\beta n(r)\mathcal{V}_{ee}(\mathbf{r}%
,\mathbf{r}^{\prime })=\delta \left( \mathbf{r}-\mathbf{r}^{\prime }\right)
-n(r)c\left( \mathbf{r},\mathbf{r}^{\prime }\right) .  \label{b.15}
\end{equation}%
It follows from the generalized Ornstein-Zernicke equation (\ref{a.15}) that
the inverse of $\epsilon \left( \mathbf{r},\mathbf{r}^{\prime };z=0\right) $
is
\begin{equation}
\epsilon ^{-1}\left( \mathbf{r},\mathbf{r}^{\prime };z=0\right) =s\left(
\mathbf{r}^{\prime },\mathbf{r}\right) ,  \label{b.16}
\end{equation}%
where the static structure factor is defined by%
\begin{equation}
s\left( \mathbf{r},\mathbf{r}^{\prime }\right) =\delta \left( \mathbf{r}-%
\mathbf{r}^{\prime }\right) +h\left( \mathbf{r},\mathbf{r}^{\prime }\right)
n\left( r^{\prime }\right) =\delta \left( \mathbf{r}-\mathbf{r}^{\prime
}\right) +h\left( \mathbf{r}^{\prime },\mathbf{r}\right) n\left( r\right) .
\label{b.17}
\end{equation}

The high frequency limit $\left( z\rightarrow \infty \right) $ of $\epsilon
\left( \mathbf{r},\mathbf{r}^{\prime };z\right) $ also has a simple form%
\begin{equation}
\epsilon \left( \mathbf{r},\mathbf{r}^{\prime };z\right) \rightarrow \delta
\left( \mathbf{r}-\mathbf{r}^{\prime }\right) +\frac{1}{z}\beta n(r)\int d%
\mathbf{v}\phi \left( v\right) \mathbf{v}\cdot \nabla _{\mathbf{r}}\mathcal{V%
}_{ee}(\mathbf{r},\mathbf{r}^{\prime })=\delta \left( \mathbf{r}-\mathbf{r}%
^{\prime }\right)  \label{b.18}
\end{equation}%
This implies no screening at asymptotically short times.

\section{Correlation functions}

\label{appD}

The Laplace transform of the correlation function (\ref{2.4}) is
\begin{equation}
\widetilde{C}_{AB}(z)=\int dxn(r)\phi \left( v\right) a(x)\widetilde{b}(x,z).
\label{c.1}
\end{equation}%
Substitution of the solution (\ref{b.12}) gives
\begin{equation}
\widetilde{C}_{AB}(z)=\int dxn(r)\phi \left( v\right) a(x)\mathcal{G}_{0}%
\left[ \overline{b}(x)-\mathbf{v}\cdot \mathbf{\nabla }_{\mathbf{r}}\int d%
\mathbf{r}^{\prime \prime }\beta \mathcal{V}_{ee}\left( \mathbf{r},\mathbf{r}%
^{\prime \prime }\right) \int d\mathbf{r}^{\prime }\epsilon ^{-1}\left(
\mathbf{r}^{\prime \prime },\mathbf{r}^{\prime };z\right) \widetilde{I}_{0}(%
\mathbf{r}^{\prime },z)\right] .  \label{c.2}
\end{equation}%
Considerable simplification occurs for the special case where $a(x)=a(%
\mathbf{r})$, i.e. it is independent of the velocity,%
\begin{eqnarray*}
\widetilde{C}_{AB}(z) &=&\int dxn(r)\phi \left( v\right) a(\mathbf{r})%
\mathcal{G}_{0}\overline{b}(x)-\Bigg[\int d\mathbf{r}a(\mathbf{r})\beta
n(r)\int d\mathbf{r}^{\prime \prime \prime }\int d\mathbf{v}\phi \left(
v\right) \mathcal{G}_{0}\mathbf{v}\cdot \mathbf{\nabla }_{\mathbf{r}}\delta
\left( \mathbf{r}-\mathbf{r}^{\prime \prime \prime }\right) \\
&&\times \int d\mathbf{r}^{\prime \prime }\mathcal{V}_{ee}\left( \mathbf{r}%
^{\prime \prime \prime },\mathbf{r}^{\prime \prime }\right) \int d\mathbf{r}%
^{\prime }\epsilon ^{-1}\left( \mathbf{r}^{\prime \prime },\mathbf{r}%
^{\prime };z\right) \widetilde{I}_{0}(\mathbf{r}^{\prime },z)\Bigg ]
\end{eqnarray*}%
\begin{eqnarray}
&=&\int dxn(r)\phi \left( v\right) a(\mathbf{r})\mathcal{G}_{0}\overline{b}%
(x)+\int d\mathbf{r}a(\mathbf{r})\int d\mathbf{r}^{\prime \prime \prime }\pi
\left( \mathbf{r},\mathbf{r}^{\prime \prime \prime };z\right)  \nonumber \\
&&\times \int d\mathbf{r}^{\prime \prime }\mathcal{V}_{ee}\left( \mathbf{r}%
^{\prime \prime \prime },\mathbf{r}^{\prime \prime }\right) \int d\mathbf{r}%
^{\prime }\epsilon ^{-1}\left( \mathbf{r}^{\prime \prime },\mathbf{r}%
^{\prime };z\right) \widetilde{I}_{0}(\mathbf{r}^{\prime },z).  \label{c.3}
\end{eqnarray}%
where use has been made of the definition (\ref{b.8}) for $\pi \left(
\mathbf{r},\mathbf{r}^{\prime \prime \prime };z\right) $. It follows from (%
\ref{b.11}) that%
\begin{equation}
\int d\mathbf{r}^{\prime \prime \prime }\pi \left( \mathbf{r},\mathbf{r}%
^{\prime \prime \prime };z\right) \int d\mathbf{r}^{\prime \prime }\mathcal{V%
}_{ee}\left( \mathbf{r}^{\prime \prime \prime },\mathbf{r}^{\prime \prime
}\right) \epsilon ^{-1}\left( \mathbf{r}^{\prime \prime },\mathbf{r}^{\prime
};z\right) =\epsilon ^{-1}\left( \mathbf{r},\mathbf{r}^{\prime };z\right)
-\delta \left( \mathbf{r}-\mathbf{r}^{\prime }\right) ,  \label{c.4}
\end{equation}%
so (\ref{c.3}) becomes%
\begin{equation}
\widetilde{C}_{AB}(z)=\int dxn(r)\phi \left( v\right) a(\mathbf{r})\mathcal{G%
}_{0}\overline{b}(x)+\int d\mathbf{r}a(\mathbf{r})\int d\mathbf{r}^{\prime }%
\left[ \epsilon ^{-1}\left( \mathbf{r},\mathbf{r}^{\prime };z\right) -\delta
\left( \mathbf{r}-\mathbf{r}^{\prime }\right) \right] \widetilde{I}_{0}(%
\mathbf{r}^{\prime },z)  \nonumber
\end{equation}%
\begin{equation}
=\int d\mathbf{r}\int d\mathbf{v}n\left( r\right) \phi \left( v\right) \int d%
\mathbf{r}^{\prime }a(\mathbf{r}^{\prime })\epsilon ^{-1}\left( \mathbf{r}%
^{\prime },\mathbf{r};z\right) \mathcal{G}_{0}\overline{b}(x)  \label{c.5}
\end{equation}%
where (\ref{b.9}) for $\widetilde{I}_{0}(\mathbf{r}^{\prime },z)$ has been
made explicit, and the dummy labels $\mathbf{r},\mathbf{r}^{\prime }$ have
been interchanged. Finally, this can be put in the simple form
\begin{equation}
\widetilde{C}_{AB}(z)=\int d\mathbf{r}d\mathbf{v}n\left( r\right) \phi
\left( v\right) a_{s}(\mathbf{r};z)\mathcal{G}_{0}\overline{b}(x),
\label{c.6}
\end{equation}%
where%
\begin{equation}
a_{s}(\mathbf{r};z)=\int d\mathbf{r}^{\prime }a(\mathbf{r}^{\prime
})\epsilon ^{-1}\left( \mathbf{r}^{\prime },\mathbf{r};z\right)  \label{c.7}
\end{equation}%
Use has been made of the fact that $n\left( r\right) \phi \left( v\right) $
commutes with $\mathcal{L}_{0}$.

\subsection{Dynamic structure factor and field autocorrelation function}

The dynamic structure factor is the autocorrelation function for the
electron density, corresponding to
\begin{equation}
a(x)=\delta \left( \mathbf{r}-\mathbf{q}\right) ,\hspace{0.25in}b(x)=\delta
\left( \mathbf{r}-\mathbf{q}^{\prime }\right) .  \label{c.8}
\end{equation}%
The correlation function (\ref{c.6}) in this case is%
\begin{equation}
\widetilde{C}(\mathbf{q},\mathbf{q}^{\prime };z)=\int d\mathbf{r}d\mathbf{v}%
n\left( r\right) \phi \left( v\right) \epsilon ^{-1}\left( \mathbf{q},%
\mathbf{r};z\right) \mathcal{G}_{0}\left( z\right) s\left( \mathbf{r},%
\mathbf{q}^{\prime }\right) ,  \label{c.9}
\end{equation}%
where $s\left( \mathbf{r},\mathbf{q}^{\prime }\right) $ is the static
structure factor of (\ref{b.17}).

The electric field autocorrelation follows from (\ref{c.9}) by integration

\begin{equation}
\widetilde{C}(z)=\int d\mathbf{q}d\mathbf{q}^{\prime }\mathbf{e}(\mathbf{q}%
)\cdot \widetilde{C}(\mathbf{q},\mathbf{q}^{\prime };z)\mathbf{e}(\mathbf{q}%
^{\prime })=\int d\mathbf{r}d\mathbf{v}n\left( r\right) \phi \left( v\right)
\mathbf{e}_{s}(\mathbf{r};z)\mathcal{G}_{0}\mathbf{e}_{s}(\mathbf{r})
\nonumber
\end{equation}%
\begin{equation}
\mathbf{e}_{s}(\mathbf{r};z)=\int d\mathbf{qe}(\mathbf{q})\epsilon
^{-1}\left( \mathbf{q},\mathbf{r};z\right) ,\hspace{0.25in}\mathbf{e}_{s}(%
\mathbf{r})=\int d\mathbf{qe}(\mathbf{q})s\left( \mathbf{r},\mathbf{q}%
\right) =\int d\mathbf{qe}(\mathbf{q})\epsilon ^{-1}\left( \mathbf{q},%
\mathbf{r};0\right)  \label{c.12}
\end{equation}


\begin{thebibliography}{99}
\bibitem{fetter} A. Fetter and J. Walecka,\emph{\ Quantum Theory of Many
Particle Systems}, (McGraw-Hill, NY, 1971).

\bibitem{hansen} J-P Hansen and I. MacDonald, \emph{Theory of Simple Liquids}%
, (Academic Press, San Diego, 1990).

\bibitem{filinov} for a recent review see A.~Filinov, V. Golubnychiy,
M.~Bonitz, W.~Ebeling, and J.~Dufty, Phys. Rev. E \textbf{70}, 046411 (2004).

\bibitem{ilya} B. Talin, A. Calisti, J. Dufty, I. Pogorelov , Phys. Rev. E
77, 036410 (2008); J. W. Dufty, I. Pogorelov, B. Talin, and A. Calisti, J.
Phys. A \textbf{36}, 6057 (2003); J. Dufty, B. Talin, and A. Calisti, in
\emph{Theory of Energy Deposition}, Adv. Quant. Chem. \textbf{46}, 293
(2004).

\bibitem{cauble} R. Cauble and D. Boercker, Phys. Rev. A \textbf{28}, 944
(1983).

\bibitem{Murillo} M. Murillo, Phys. Plasmas \textbf{5}, 3116 (1998)

\bibitem{Alastuey} A. Alastuey, J. Stat. Phys. \textbf{48}, 839 (1987).

\bibitem{Stambulchik} E. Stambulchik, D.V. Fisher, Y. Maron, H.R. Griem, and
S. Alexiou, High Energy Density Physics 3, 272 (2007).

\bibitem{wrighton} J. Wrighton, Ph.D. thesis, University of Florida, 2004.

\bibitem{ebeling} W. Ebeling, A. Filinov, M. Bonitz, V. Filinov, and T.
Pohl, J. Phys. A: Math. Gen. \textbf{39}, 4309 (2006).

\bibitem{dielectric} Strictly speaking the dielectric function is defined in
terms of the response function and is different in general from that given
here. They agree only in the weak electron - electron limit. However, it is
a convenient terminology as the function considered here does determine the
collective modes.

\bibitem{lewis} M. Lewis, Phys. Rev. \textbf{121}, 501 (1964); E. Dufour, A.
Calisti, B. Talin, M. Gigosos, M. Gonz\'{a}lez, T. del R\'{\i}o
Gaztelurrutia, and J. Dufty, Phys. Rev. E \textbf{71}, 066409 (2005).

\bibitem{Gunnarson} O. Gunnarson, M. Jonson, and B. Lundquist, Solid State
Commun. \textbf{24,} 765 (1977); M. Murillo and J. Weisheit, Phys. Reports
302, 1 (1998).
\end{thebibliography}
\end{document}